\documentclass[a4paper,11pt]{article}
\pdfoutput=1 

\usepackage{jcappub} 

\usepackage[T1]{fontenc} 

\usepackage{float} 

\usepackage{caption}
\usepackage{subcaption}

\newcommand{\newc}{\newcommand}
\newc{\gsim}{\lower.7ex\hbox{$\;\stackrel{\textstyle>}{\sim}\;$}}
\newc{\lsim}{\lower.7ex\hbox{$\;\stackrel{\textstyle<}{\sim}\;$}}
\newc{\gev}{\,{\rm GeV}}
\newc{\mev}{\,{\rm MeV}}
\newc{\ev}{\,{\rm eV}}
\newc{\kev}{\,{\rm keV}}
\newc{\tev}{\,{\rm TeV}}

\newc{\mz}{M_Z}
\newc{\mw}{m_{\rm weak}}
\newc{\nr}[1]{N^c_R{}_{#1}}
\usepackage{amsmath}

\def\beq{\begin{equation}}
\def\eeq{\end{equation}}
\def\bea{\begin{eqnarray}}
\def\eea{\end{eqnarray}}
\def\bitem{\begin{itemize}}
\def\eitem{\end{itemize}}
\newcommand{\bec}{\begin{center}}
\newcommand{\eec}{\end{center}}
          \newcommand{\cm}{{\mathrm {cm}}}

\newcommand{\half}{\frac{1}{2}}

  \newcommand{\GeV}{{\mathrm {GeV}}}
   \newcommand{\TeV}{{\mathrm {TeV}}}

\def\bar#1{\overline{#1}}

\def\inv{^{\raise.15ex\hbox{${\scriptscriptstyle -}$}\kern-.05em 1}}
\def\lbar{{\lower.35ex\hbox{$\mathchar'26$}\mkern-10mu\lambda}} 

\let\<=\langle
\let\>=\rangle

\let\+=\uparrow

\def\al{\alpha} 
\def\be{\beta}
\def\ga{\gamma}

\def\ep{\epsilon}

\def\ka{\kappa}

\def\si{\sigma}

\def\Ga{\Gamma}

\def\La{\Lambda}

\def\Om{\Omega}

\newcommand{\ben}{\begin{equation}}
\newcommand{\een}{\end{equation}}
\newcommand{\ba}{\begin{array}}
\newcommand{\ea}{\end{array}}
\newcommand{\bit}{\begin{itemize}}
\newcommand{\eit}{\end{itemize}}

\newcommand{\dm}{dark matter}			
\newcommand{\Mpl}{M_{\rm Pl}}			
\newcommand{\mpl}{m_{\rm Pl}}			
\newcommand{\mchi}{m_{\chi}}			
\newcommand{\nchi}{n_{\chi}}				
\newcommand{\NXchi}{N_{\chi}}			
\newcommand{\nchieq}{n_{\chi, \textrm{eq}}}	
\newcommand{\mX}{m_{X}}				
\newcommand{\fX}{f_{X}}					
\newcommand{\EbarX}{\bar{E}_{X}}			
\newcommand{\sichi}{\sigma_{\chi}}			
\newcommand{\Ychi}{Y_{\chi}}				
\newcommand{\Ychieq}{Y_{\chi, \textrm{eq}}}	
\newcommand{\vd}{v_{\textrm{d}}}			
\newcommand{\rhod}{\rho_{\textrm{d}}}		
\newcommand{\eosd}{w_{\textrm{d}}}			
\newcommand{\Omd}{\Om_{\textrm{d}}}		

\newcommand{\refsub}{\chi}				
\newcommand{\Qref}{Q_\refsub}				
\newcommand{\tref}{t_\refsub}	                            
\newcommand{\Tref}{T_\refsub}				
\newcommand{\rhoref}{\rho_\refsub}			
\newcommand{\Href}{H_\refsub}				
\newcommand{\xref}{x_\refsub}				
\newcommand{\Dref}{D_\refsub}			

\newcommand{\geff}{g_*}					
\newcommand{\heff}{h_*}					

\newcommand{\Xprodpar}{q_X}			
\newcommand{\chimult}{r_\chi}			 	

\newcommand{\Tfree}{T_{\textrm{fr}}}		
\newcommand{\xifree}{\xi_\text{fr}}			

\newcommand{\Qcusp}{Q_\text{c}}
\newcommand{\FTscen}{FT}
\newcommand{\CAscen}{CE}
\newcommand{\TD}{TD}
\newcommand{\muPar}{\ep}

\newcommand{\tCE}{t_\text{ce}}
\newcommand{\ellCE}{\ell_\text{ce}}
\newcommand{\TCE}{T_\text{ce}}
\newcommand{\beCE}{\beta_\text{ce}}
\newcommand{\Td}{T_\text{d}}
\newcommand{\xd}{x_\text{d}}

\newcommand{\lcdm}{$\La$CDM}
\newcommand{\paramFT}{P_\text{FT}}
\newcommand{\paramCE}{P_\text{CE}}

\title{Dark Matter from Decaying Topological Defects}

\author[a,b]{Mark~Hindmarsh}
\author[c]{Russell~Kirk}
\author[c,d]{Stephen M. West}

\affiliation[a] {Dept.\ of Physics and Astronomy, University of Sussex, 
Brighton BN1 9QH, U.K.}
\affiliation[b]{Helsinki Institute of Physics, P.O.\ Box 64, 00014 Helsinki University, Finland} 
\affiliation[c]{Dept.\ of Physics,
Royal Holloway University of London, Egham, Surrey TW20 0EX, U.K.}
\affiliation[d]{Rutherford Appleton Laboratory, Chilton, Didcot, OX11 0QX, UK} 
\emailAdd{m.b.hindmarsh@sussex.ac.uk}
\emailAdd{russell.kirk.2008@live.rhul.ac.uk}
\emailAdd{stephen.west@rhul.ac.uk}

\abstract{
We study dark matter production by decaying topological defects, in particular cosmic strings.  In topological defect or ``top-down'' (\TD) scenarios, the dark matter injection rate varies as a power law with time with exponent $p-4$. 
We find a formula in closed form for the yield for all $p< 3/2$, which accurately reproduces the solution of the Boltzmann equation. 
We investigate two scenarios ($p=1$, $p=7/6$) motivated by cosmic strings which decay into TeV-scale states with a high branching fraction into dark matter particles. 
For dark matter models annihilating either by s-wave or p-wave,  we find the regions of parameter space where the \TD\ model can account for the dark matter relic density as measured by Planck. We find that topological defects can be the principal source of dark matter, even when the standard freeze-out calculation under-predicts the relic density and hence can lead to  potentially large ``boost factor'' enhancements in the dark matter annihilation rate. We examine dark matter model-independent limits on this scenario arising from unitarity and discuss example model-dependent limits coming from indirect dark matter search experiments.
In the four cases studied, the upper bound on $G\mu$ for strings with an appreciable channel into TeV-scale states is significantly more stringent than the current Cosmic Microwave Background limits.
}


\begin{document}
\maketitle
\flushbottom

\section{Introduction}

There is extremely strong indirect evidence that the universe has a substantial component of energy density in the form of very weakly interacting non-relativistic particles - dark matter.  The most recent measurement of the dark matter relic abundance by Planck gives $\Om_{\rm DM}h^2 = 0.1186\pm 0.0031$ in the canonical \lcdm\ model \cite{Ade:2013ktc}. The theoretical picture of dark matter invokes beyond the standard model (BSM) physics commonly in the form of an electrically neutral weakly interacting massive particle (WIMP) that undergoes freeze-out \cite{freeze-out1, freeze-out2, freeze-out3} in the Early Universe. 

In freeze-out the relic abundance of dark matter is in general determined by the dark matter annihilation rate to standard model states. As a consequence the requirements placed on the annihilation rate to get the correct abundance can be cross-correlated directly with limits from direct and indirect dark matter search experiments. This link can provide serious constraints on dark matter models due to the almost one to one correspondence between the required dark matter annihilation rate for freeze-out and the annihilation rate of dark matter now. 

There exists a range of exceptions to this correspondence that serve to loosen the tight link between the physics of freeze-out in the Early Universe and that of dark matter search experiments taking data now. The well known examples of co-annihilation, near mass threshold annihilation and resonant annihilation can change the freeze-out dynamics \cite{Griest:1990kh}. This link can also be broken by considering alternatives to the freeze-out generation mechanism. For example, the dark matter abundance may be generated by the freeze-in mechanism \cite{Hall:2009bx, Williams:2012pz} or non-thermally by decays of gravitinos (for examples of such models see 
\cite{Moroi:1999zb,Hindmarsh:2013xg} and for discussions of their cosmological implications see \cite{Hisano:2000dz,Lin:2000qq}).  
For a review of other dark matter candidates and production mechanisms see, for example, \cite{Steffen:2008qp}.

BSM physics often also predicts phase transitions at which topological defects are formed \cite{Kibble:1976sj}.  If the broken symmetry is a gauge symmetry, the only cosmologically acceptable defects are cosmic strings \cite{Hindmarsh:2011qj}. 
Indeed, in scenarios combining hybrid inflation with supersymmetric grand unification, cosmic string formation is generic \cite{Jeannerot:2003qv}. 

The strings decay into the particles of the fields from which they are made, and into gravitational radiation, with uncertain branching fractions \cite{Hindmarsh:2011qj}. If the particles are coupled to dark matter particles, there will be another source of dark matter production from the decaying topological defects. This additional source of dark matter particles could then modify the relationship between the annihilation rate required for successful freeze-out and the annihilation rate of dark matter today. 

Previous work on dark matter from decaying strings \cite{Jeannerot:1999yn} considered a particular production mechanism: dark matter particles were presumed to be produced, in small numbers, only when loops of string had shrunk to radii the same order as the string width.

In this paper, rather than restricting ourselves to a particular scenario, we envisage a generic top-down 
(\TD) production mechanism 
for dark matter particles, analogous to the \TD\ cosmic ray production scenario
\cite{Bhattacharjee:1998qc}.
In the \TD\ dark matter production scenario, the usual Boltzmann equation 
is supplemented with a source of dark matter particles having a 
power-law dependence on time.  The dark matter is envisaged to result from the decays of a new sector of particles denoted $X$, whose lifetime is assumed to be $\sim 1/m_X$ and well below $H^{-1}$ when $T\sim m_X$.
We parametrise the source by the energy density injection rate $\Qref$ when the temperature is equal to the mass of the dark matter particle $\mchi$, the exponent of the power law $p$, the average energy of the particles produced by the decay $\EbarX$, and the average multiplicity of dark matter particles by the subsequent decay of the particles 
$\NXchi$. 
We find numerical solutions for the dark matter yield as a function of scale factor, and a closed-form formula for its asymptotic value.

We study the constraints on cosmic strings in two particle production scenarios.
In the first (which we denote \FTscen\ as it is motivated by direct simulation of a field theory \cite{Vincent:1997cx,Bevis:2006mj,Hindmarsh:2008dw}), the strings decay entirely into propagating modes of the fields from which they are made, with a high branching fraction into $X$ particles.
In the second, the branching fraction into $X$ particles is small, with the primary decay channel 
being gravitational radiation from long-lived oscillating loops \cite{Damour:2004kw}. 
Here, the most important source of $X$ particles is thought to be emission from cusps (where a length of string doubles back on itself), and we denote this scenario \CAscen.  
We assume that the masses of the relevant $X$ particles (i.e.\ those which couple to the dark matter) are TeV-scale \cite{Bhattacharjee:1997in}.

In each scenario, we also allow either the s-wave or the p-wave dark matter annihilation channel to dominate, giving four models in all. We find the regions of parameter space where production from string decay can account for the Planck value of the dark matter relic density even when the standard freeze-out calculation would otherwise under-predict it. Moreover, we highlight that by increasing the contribution from cosmic strings the dark matter annihilation rate can be increased whilst still maintaining the correct relic abundance. This leads to the possibility of having a large enhancement in the dark matter annihilation rate over what would be expected from standard freeze-out. As a result this scenario may be a viable way to produce ``boost factors'' that can enhance the size of indirect dark matter signals. A model-independent limit on this enhancement comes from the unitarity of the dark matter annihilation cross section. We use this unitarity limit to constrain the string parameters, in particular we derive interesting bounds on the string tension parameter $G\mu$. In all cases studied, the bounds on $G\mu$ (displayed in Table \ref{vdlims}) are well below the current Cosmic Microwave Background limits
$G\mu \lesssim 3.2 \times10^{-7}$ (FT) and $G\mu \lesssim 1.3 \times10^{-7}$ (CE) \cite{Ade:2013xla}, which 
correspond to the Grand Unified or inflation scale. The complete set of our results and constraints, including an example of more model-dependent limits coming from indirect searches for dark matter are presented in Section~\ref{cons}.

\section{Top-down models of dark matter production}

In \TD\  models of particle production in the early universe \cite{Bhattacharjee:1998qc}, 
there is an exotic component (perhaps topological defects like cosmic strings) which can decay into
a new sector of massive particles $X$, which in turn decay into dark matter.  The $X$ particles may also decay into particles of the Standard Model, giving rise to high-energy cosmic and $\ga$-rays. 

We denote the average energy density of the \TD\ source by $\rhod$, and its density fraction $\Omd = \rhod/\rho$, where $\rho$ is the total energy density. Denoting the average equation of state parameter $\eosd$, we find that the total energy density injection rate into other species from \TD\ decay is 
\ben
Q = 3(w-\eosd)\Omd\rho H,
\een
where $\rho$ is the total energy density, $w$ its average equation of state parameter, and $H$ the Hubble parameter. 
We assume that a fraction $\fX$ of the energy from defect decay results in the production of new states labelled $X$, with masses $\mX$, and average energy $\EbarX$.  
The decays of these particles are assumed to produce on average $\NXchi$ dark matter particles $\chi$. 

With the above definitions, the dark matter particle number density injection rate is 
\ben
j_\chi^{\rm inj} = \fX\NXchi\frac{Q}{\EbarX}. 
\een
Similarly to \cite{Bhattacharjee:1998qc}, in a \TD\ model the energy density injection rate $Q$ is supposed to depend on time as  
\beq
\label{eq:eninj}
Q(t)=\Qref\left(\frac{t}{\tref}\right)^{p-4},
\eeq
where $\tref$ is a reference time in the radiation-dominated era, which will be chose to be when the temperature is equal to the mass of the \dm\ particle.

We will explore two specific \TD\ scenarios, in both of which the source is provided by cosmic strings \cite{Kibble:1976sj, VilShe94, Hindmarsh:2011qj}. The string forms at a phase transition at an energy scale $\vd$, which is taken to be larger than $m_X$. Both scenarios are described in detail in Appendix \ref{s:EvoStr}, and here we merely summarise the important features.

In the first scenario, motivated by direct numerical simulation of strings in the Abelian Higgs field theory \cite{Vincent:1997cx,Bevis:2006mj,Hindmarsh:2008dw}, strings decay into $X$ particles with a branching fraction $\fX \simeq 1$. In this case (which we shall denote \FTscen) the power law index takes the value $p=1$.  In the second scenario, the dominant decay mode is gravitational radiation emitted by a population of loops oscillating according the the Nambu-Goto equations of motion \cite{Vilenkin:1981bx}. There are also subdominant modes of particle production, of which the most important at late times is cusp emission \cite{Brandenberger:1986vj,BlancoPillado:1998bv,Vachaspati:2009kq}.  
This mechanism gives rise to a power law index $p=7/6$ in the radiation era. 
We will denote this second scenario \CAscen. 
Particle emission can also take place at the last stages of collapse of loops \cite{Jeannerot:1999yn}. It results in injection with $p = -1/2$ and so is subdominant at late times.

There are two main cases to study for the values of $\NXchi$ and $\EbarX$. The first is where a few $X$ states with $\mX \sim 10^{15}$\;GeV are produced with a large average energy, $\EbarX \sim 10^{15}$\;GeV. The number of dark matter particles produced by $X$ decays is model-dependent. We will not consider this scenario further in this paper.

We will instead study a second scenario for $X$ production, where many lower mass states are produced with low average energies.  We shall consider the case $\bar{E}_X\sim10^{3}$ GeV, 
appropriate for topological defects produced in supersymmetric theories where the vacuum expectation value $\vd$ is in a (combination of) fields in a flat direction \cite{Bhattacharjee:1997in} or to strings with condensates of light fields \cite{Vachaspati:2009kq}.  We assume that each of the $X$ states decays to a low number of dark matter particles, likely to be even due to the symmetry stabilising them. However, we will conservatively take $N_{\chi} = 1$. 

\section{Dark matter, Boltzmann equation with sources}

The Boltzmann equation for the number density, $\nchi$, of dark matter states reads
\beq
\label{eq:be1}
\dot{n}_{\chi}+3H\nchi =-\langle \sichi v \rangle \left(\nchi^2- \nchieq^2\right)+\frac{\NXchi\fX Q(t)}{\EbarX},
\eeq
where $\langle \sichi v \rangle$ is the thermally-averaged dark matter annihilation cross section, with $v$ the relative velocity of the annihilating dark matter states, and $\nchieq$ is the equilibrium dark matter number density. The second term on the right hand side accounts for the production of dark matter states from defect decays, where we have assumed that the decay of $X$ states into dark matter is instantaneous, or at least occurs with a rate that 
is much larger than the dark matter annihilation rate. The dark matter states, highly non-thermal when produced, are assumed to rapidly approach kinetic equilibrium.  This is a reasonable assumption given that the scattering rate of the dark matter particles on the thermal bath is high compared with the Hubble rate at the freeze-out temperatures we consider, $T \sim 20\,\GeV$, unless the scattering cross-section is orders of magnitude smaller than the annihilation cross-section. 

As usual, it is convenient to change variable $x=\mchi/T$, and expand the dark matter annihilation cross section in powers of $x$, $\langle \si_{\chi} v \rangle=\si_0 x^{-n}$. A dominant s-wave annihilation process corresponds to $n=0$, and $n=1$ to a dominant p-wave annihilation. For simplicity, we will assume that the effective numbers of relativistic degrees of freedom contributing to the energy and entropy densities ($\geff$ and $\heff$) are equal and slowly varying during the relevant times.

The Boltzmann equation \eqref{eq:be1} can be rewritten in terms of the dark matter yield, $\Ychi=\nchi/s$, where $s$ is the entropy density, as
\beq
\label{eq:Boltz}
\frac{d \Ychi}{dx}=-\frac{A}{x^{n+2}} \left(\Ychi^2-\Ychieq^2\right) + \frac{B}{x^{4-2p}},
\eeq
where
\beq
A=\sqrt{\frac{\pi}{45}}\sqrt{g_*}\Mpl\mchi\si_0, \quad
B=\frac{3}{4}\left(\frac{\NXchi\mchi}{\EbarX}\right)\left(\frac{\Qref\fX}{\rhoref \Href}\right).
\eeq
Here,  $\rho_\refsub$ and $H_\refsub$ are the energy density and the Hubble parameter evaluated at the reference temperature $T=\mchi$ or $x = 1$, while $\Mpl = 1/\sqrt{G} \simeq 1.22\times10^{19}\;\mbox{GeV}$ is the Planck mass.

We will be particularly interested in the constraints on the $X$ injection rate parameter
\ben
\Xprodpar = \frac{\Qref f_X}{\rhoref\Href}
\label{e:XProParDef}
\een
in the above scenarios.  It is clear that they will depend on assumptions made about the $\chi$ multiplicity parameter
\ben
\label{e:chimultDef}
\chimult = \frac{\NXchi\mchi}{\EbarX}.
\een

\section{Numerical solution} \label{sec:Numeric}

\begin{figure}[t]
\begin{subfigure}{.5\textwidth}
  \hspace{-1cm}
  \includegraphics[width=8cm]{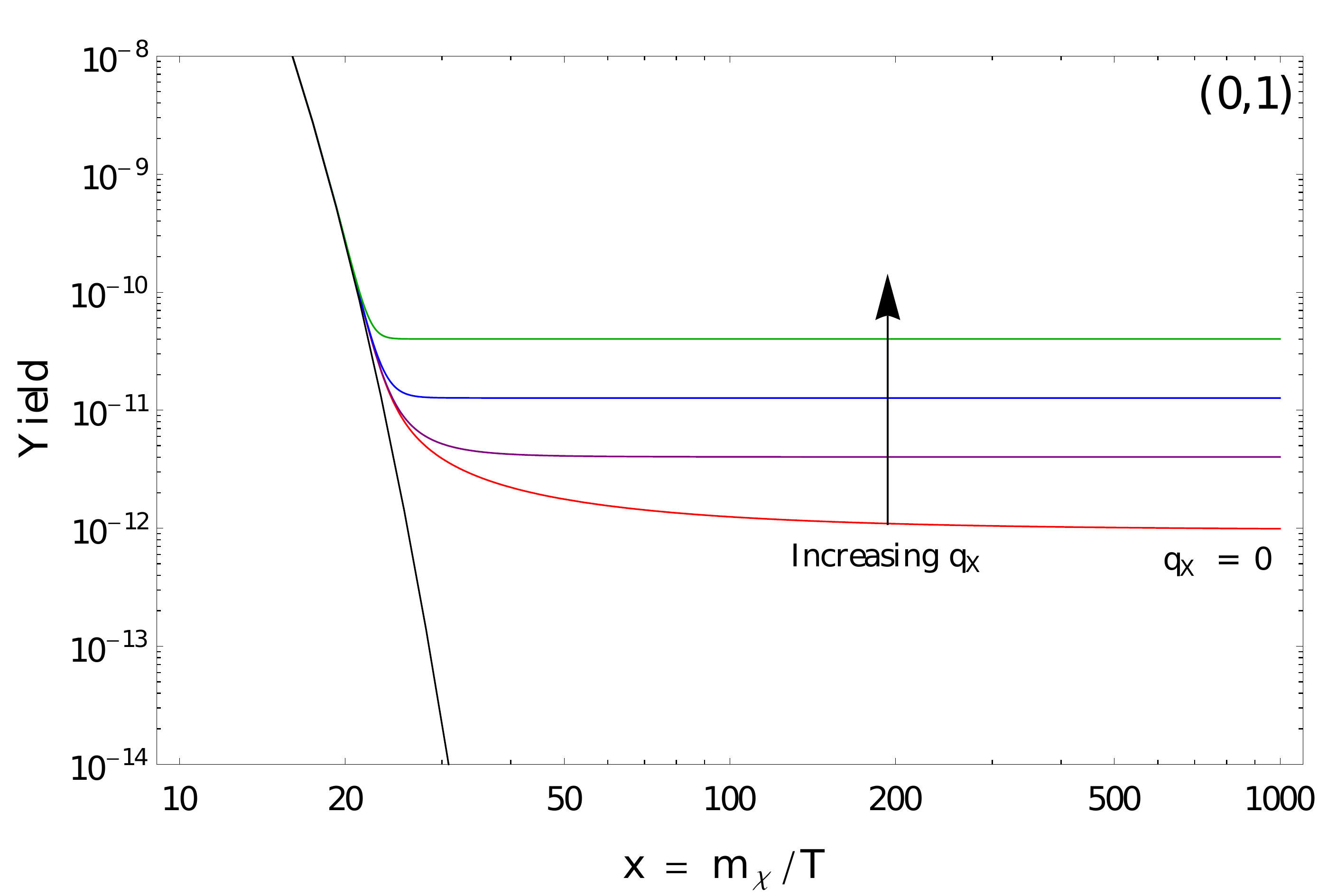}
  \caption{\hspace{0.2cm}\phantom{    .}}
  \label{fig:YvXN0P1}
\end{subfigure}%
\begin{subfigure}{.5\textwidth}
  \includegraphics[width=8cm]{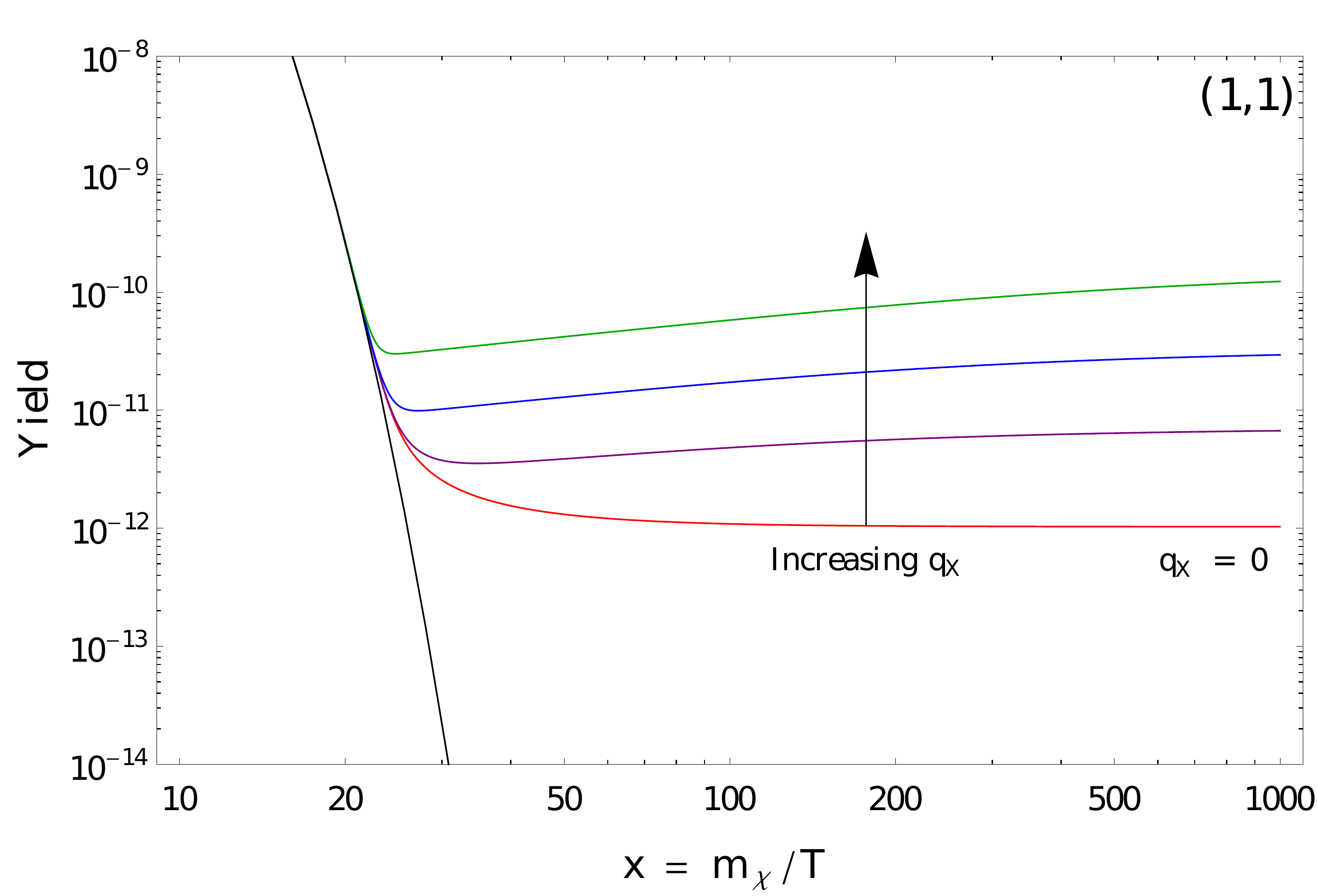}
  \caption{\hspace{-1.7cm}\phantom{a}}
  \label{fig:YvXN1P1}
\end{subfigure}%
\\\begin{subfigure}{.5\textwidth}
  \hspace{-1cm}
  \includegraphics[width=8cm]{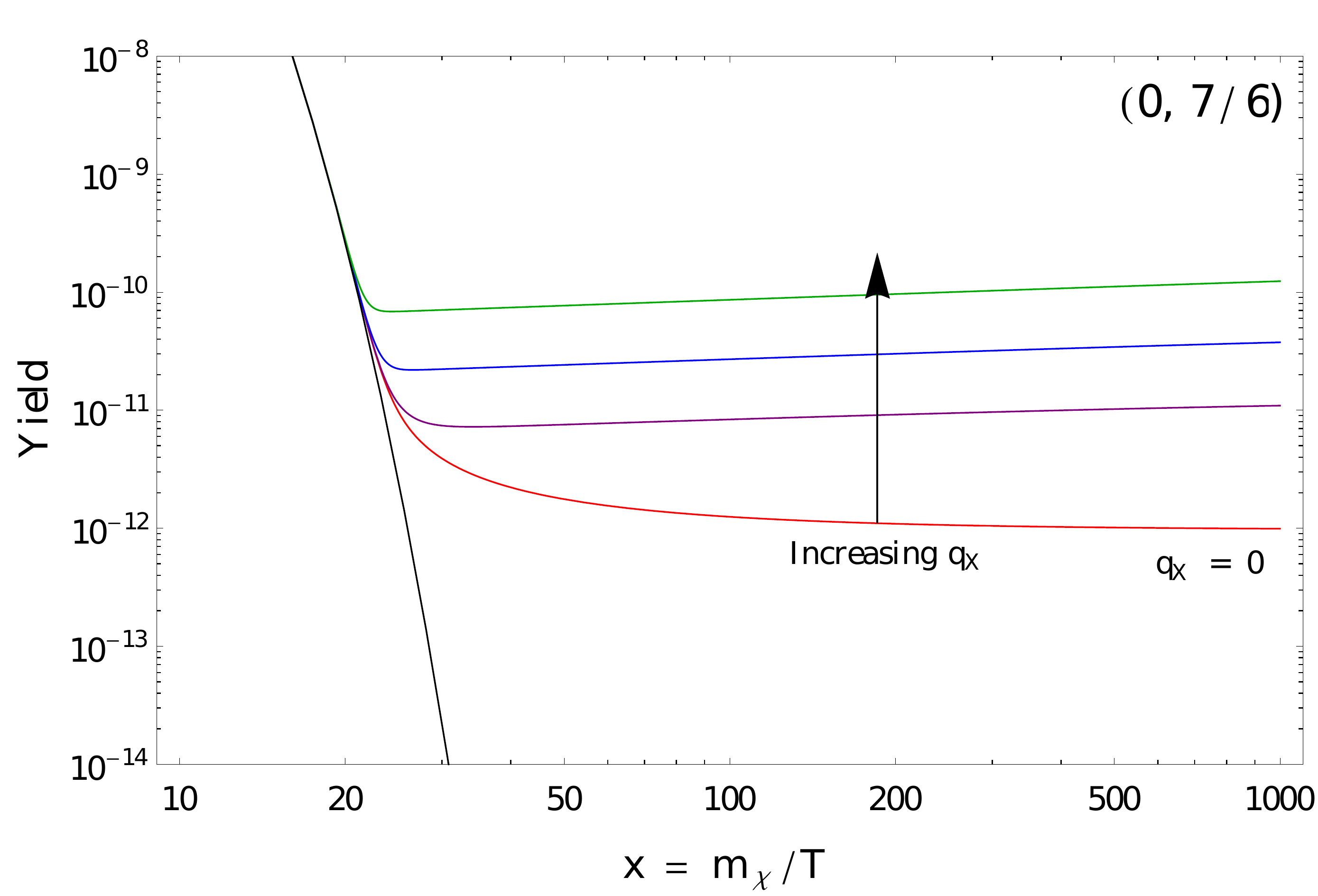}
  \caption{\hspace{0.2cm}\phantom{    .}}
  \label{fig:YvXN0P76}
\end{subfigure}%
\begin{subfigure}{.5\textwidth}
  \includegraphics[width=8cm]{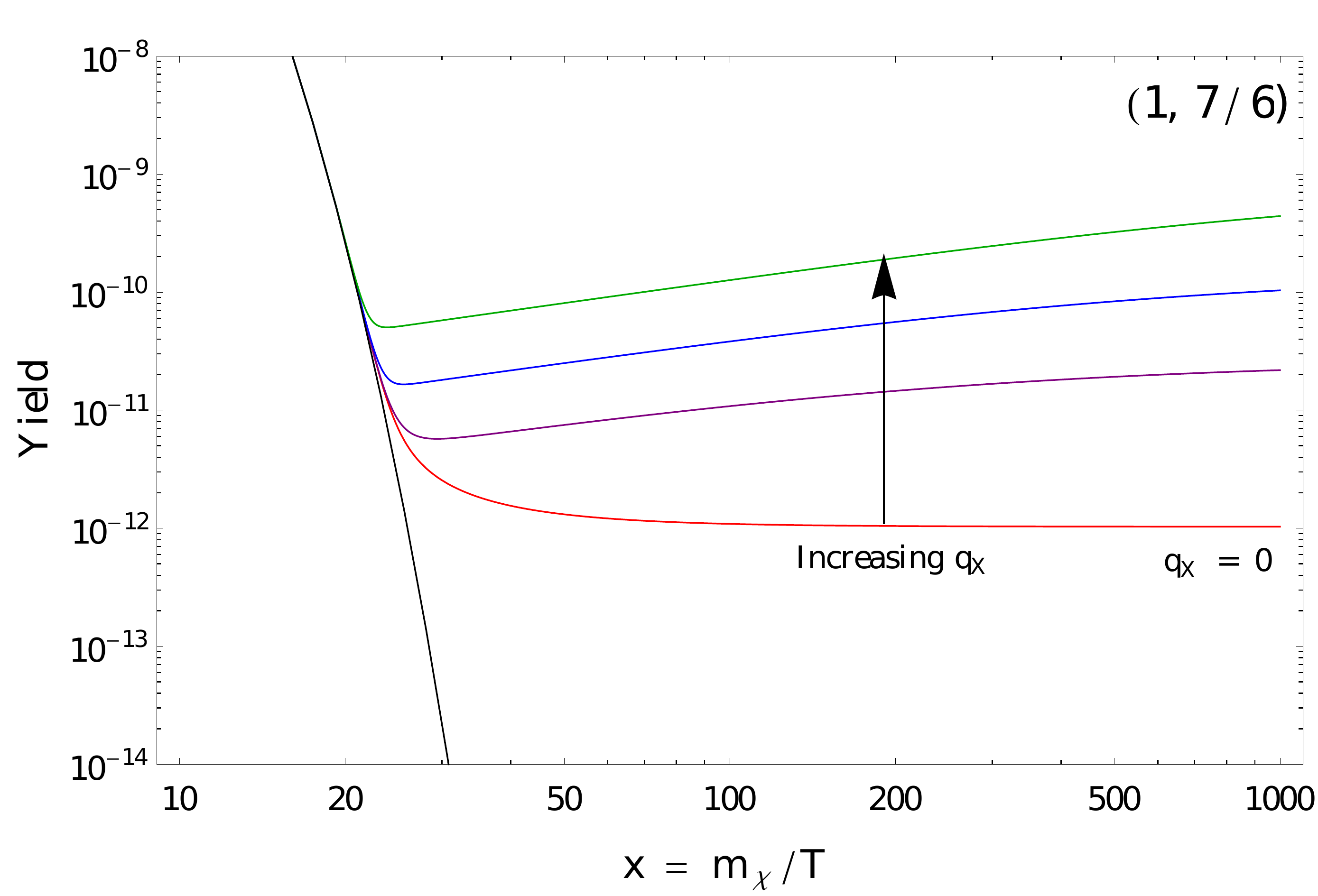}
  \caption{\hspace{-1.7cm}\phantom{a}}
  \label{fig:YvXN1P76}
\end{subfigure}%
\caption{Numerical evaluation of the Boltzmann equation \eqref{eq:Boltz} where we have set $m_{\chi}=500$\;GeV, $\chimult=0.5$, and for (\subref{fig:YvXN0P1}) $(n, p)=(0, 1)$ and $\si_0=1.6\times 10^{-26}\;$cm$^3$s$^{-1}$ (\subref{fig:YvXN1P1}) $(n, p)=(1, 1)$ and $\si_0=7\times 10^{-25}\;$cm$^3$s$^{-1}$  (\subref{fig:YvXN0P76}) $(n, p)=(0, 7/6)$ and $\si_0=1.6\times 10^{-26}\;$cm$^3$s$^{-1}$ and (\subref{fig:YvXN1P76}) $(n, p)=(1, 7/6)$ and $\si_0=7\times 10^{-25}\;$cm$^3$s$^{-1}$. For each plot, curves for different values of $\Xprodpar = 0, 10^{-9}, 10^{-8}, 10^{-7}$ are plotted. The bottom curve in each plot represents the standard freeze-out scenario with $\Xprodpar=0$. Also plotted is the equilibrium yield line represented by the solid black line.}
\label{fig:yield}
\end{figure}

In this section the Boltzmann equation \eqref{eq:Boltz} is solved numerically, leaving an approximate analytic treatment to Section~\ref{sec:Analytic}. Numerical solutions for the yield of dark matter as a function of temperature, $T$, are plotted for our four different combinations of $n$ and $p$. These values of $(n, p)$ are labelled on each plot and for Figures~\ref{fig:YvXN0P1}, \ref{fig:YvXN1P1}, \ref{fig:YvXN0P76} and \ref{fig:YvXN1P76} are $(0, 1), (1, 1), (0,7/6)$ and, $(1, 7/6)$ respectively.

In each plot the resulting yield for three values of the $X$ injection rate parameter ($\Xprodpar=10^{-9}, 10^{-8}, 10^{-7}$) are compared to the case of standard freeze-out, that is with $\Xprodpar=0$. In all plots in Figure~\ref{fig:yield} we have set $m_{\chi}=500$\;GeV, our default value of $\EbarX$ in our featured decay scenario is $10^3$ GeV, and so $\chimult = 0.5$. We also set the number of relativistic degrees of freedom to $g_* =100$.

In Figure~\ref{fig:YvXN0P1} and \ref{fig:YvXN0P76} we have chosen an annihilation process that is dominantly s-wave and therefore set $n=0$. The dark matter annihilation cross section parameter $\si_0$ has been fixed to $\si_0=1.6\times 10^{-26}\;$cm$^3$s$^{-1}$, which corresponds to the standard freeze-out value needed to generate the measured abundance for a dark matter state with a mass of $500\gev$ . 
For Figure~\ref{fig:YvXN1P1} and \ref{fig:YvXN1P76} we show the plots for $n=1$ with $\si_0=7\times 10^{-25}\;$cm$^3$s$^{-1}$, where $\si_0$ has again be chosen to give the standard freeze-out result but this time for a dominantly p-wave annihilation process. 

The plots in Figure~\ref{fig:yield} demonstrate two important effects. The first is the increase in the predicted yield with an increasing contribution from the defect decays. This result is of course expected as we would anticipate that if the production of dark matter states from the source increases the number density of dark matter states should also increase. 

The second important effect is that the post freeze-out behaviour of the yield as a function of $T$ can be linked to the relative sizes of the inverse powers of $x$ in the annihilation ($n+2$) and source ($4-2p$) terms in \eqref{eq:Boltz}. In Figure~\ref{fig:YvXN0P1}, we have that $n+2=4-2p$, and for Figures~\ref{fig:YvXN1P1}, \ref{fig:YvXN0P76} and \ref{fig:YvXN1P76}, $n+2>4-2p$. 

For Figure~\ref{fig:YvXN0P1} with $n+2=4-2p$ we see that the post-freeze-out yield becomes constant in temperature quickly after decoupling from the equilibrium yield. The reason for this is that as the temperature drops below the mass of the dark matter states the thermal bath particles (that is, those to which the dark matter annihilates) no longer have the energy to produce two dark matter states. In the absence of a source term, as in standard freeze-out, the dark matter yield will follow the equilibrium yield until the annihilation rate drops below the Hubble expansion rate and freezes-out. 

With a source term this process is modified. Now as the temperature drops below the dark matter mass we still have a source of dark matter states. This results in an increase in the yield above the equilibrium value. The yield is still decreasing, however, and at some temperature (defined as $\xd$) the first term on the right hand side of~\eqref{eq:Boltz} (the annihilation term) will be equal in magnitude to the second term (the source term). At this temperature $dY/dx=0$ and the value of the yield will be fixed at a constant value, $Y\sim\sqrt{B/A}$ and will remain at this value. The larger the value of $\Xprodpar$ the earlier this equality is reached and the lower the value of $\xd$ (see Section \ref{sec:Analytic} for the analytic treatment of this effect) and as a result the final yield will be larger. 

For Figures~\ref{fig:YvXN1P1}, \ref{fig:YvXN0P76} and \ref{fig:YvXN1P76} we have that $n+2>4-2p$ and as a result the source term will dominate (provided $\Xprodpar\neq 0$) post-freeze-out due to the annihilation rate dropping faster than the injection rate as the temperature decreases. Hence, we would expect to see an increase in the yield after annihilations are no longer efficient as dark matter states are still being produced by the decay of the topological defects. 

\begin{figure}[t]
\centerline{\includegraphics[width=9cm]{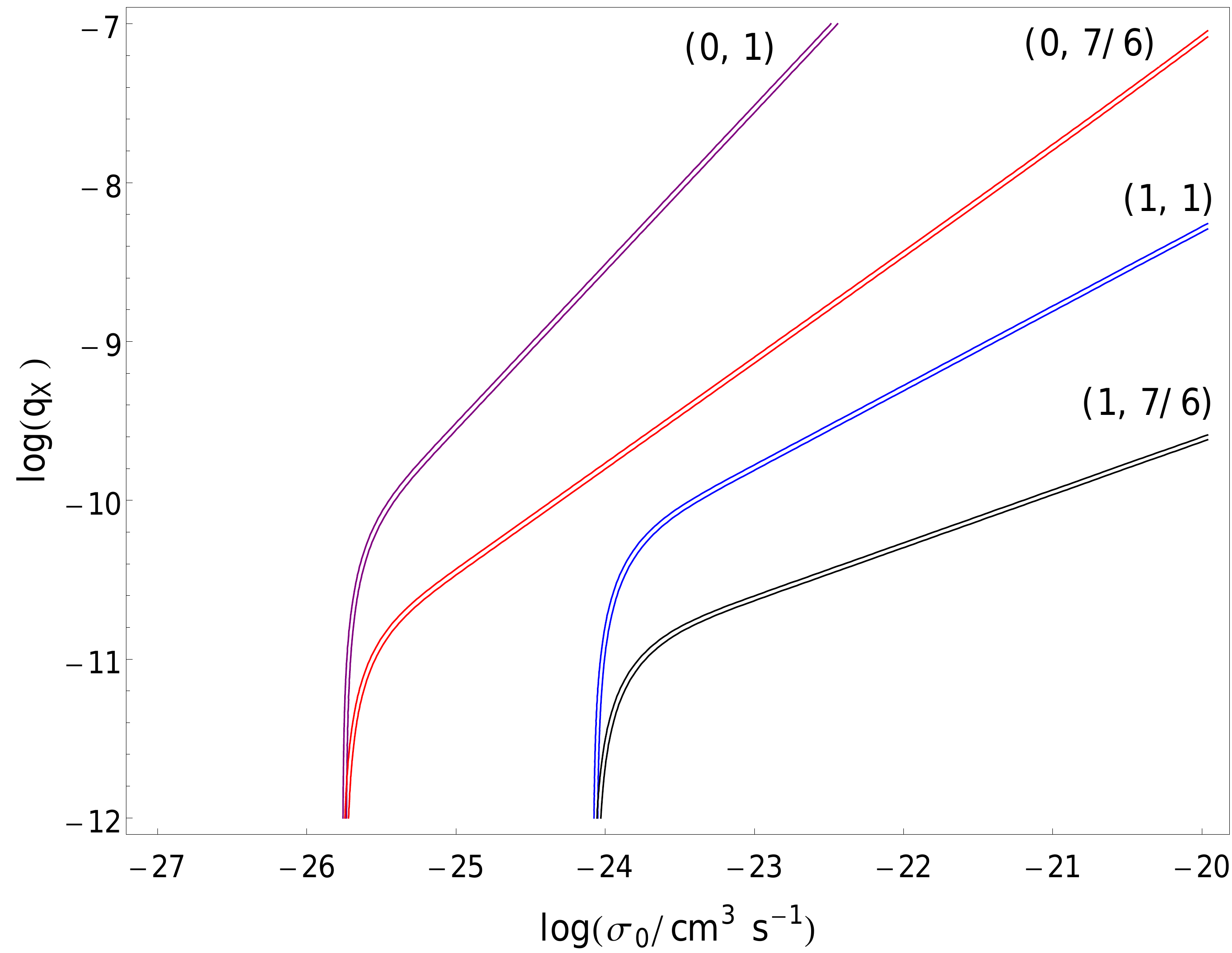}}
\caption{Contours of constant yield in the $\Xprodpar$ vs $\si_0$ parameter space set to 1$\si$ either side of the Planck-derived value of the dark matter abundance. We have used $m_{\chi}=500$ GeV, $\chimult=0.5$. Each curve is labelled by its $(n, p)$ values where $n$ is the angular momentum of the annihilation channel and $p$ is defined in (\ref{eq:eninj}).}
\label{fig:qxsig}
\end{figure}

We are of course interested in the final yield, which will determine the final abundance of the dark matter states. We can get an idea for the way in which the source term can affect this quantity from the plots in Figure~\ref{fig:yield} but a more instructional way to examine the effect of the source term is to look in the $(\Xprodpar, \si_0)$ parameter plane for our chosen $(n, p)$ values. The results are plotted in Figure~\ref{fig:qxsig}, with the curves representing contours of constant final dark matter abundance set to the value determined by Planck. We show two contours for each $(n, p)$ value that correspond to the range of abundances that are within 1$\si$ of the measured value of $\Om_{\chi}h^2$ \cite{Ade:2013ktc}. 

In Figure~\ref{fig:qxsig} for each scenario two distinct behaviours are evident. For large $\Xprodpar$ the contour adopts a straight line on the logarithmic plot suggesting that even in the region where the contribution from the decaying defects is large the final yield is still dependent on the dark matter annihilation cross section. Moving to smaller values of $\Xprodpar$, the contribution of the defect decays becomes less important with the annihilation rate becoming the dominant  process setting the abundance. The near vertical part of all curves corresponds to the scenario where we are approaching standard freeze-out. This conclusion is backed up by the fact that there are two distinct values for $\si_0$ in the small $\Xprodpar$ limit. These two values correspond to standard freeze-out via dominant s-wave (smallest value) and dominant p-wave (largest value) annihilation. 

A further effect depicted in Figure~\ref{fig:qxsig} is that for a given value of $n$ the smaller the $p$ the larger the gradient of line. This is due to the fact that the source term in~\eqref{eq:Boltz} scales as $x^{-(4-2p)}$. For larger $(4-2p)$ values (keeping all other parameter constant) the effect of this term at freeze-out is reduced. As a result, the overall size of the source term needs to be increased (that is, a larger value of $\Xprodpar$) to generate the same abundance as for smaller values of $(4-2p)$. 

An interesting point to note is that Figure~\ref{fig:qxsig} shows us that we can potentially increase the size of the annihilation cross section for a dark matter candidate but still generate the correct relic abundance. If we increased the cross section without the contribution from the source then we will under produce the dark matter during freeze-out. An immediate result is that we could expect to have larger annihilation cross sections for dark matter now and this effect could be a source of so-called particle physics boost factors in indirect detection\footnote{There are a number of different types of particle physics boost factors for indirect detection. For example the ``Sommerfeld effect'' (see, e.g. \cite{MarchRussell:2008tu}).}. We return to this point in Section~\ref{cons}.

\section{Analytic solution(s)} \label{sec:Analytic}

To get a better understanding of the dependences of the yield on the source parameters we can find, following an analogous procedure to that outlined in \cite{Kolb:1994}, an approximate analytical solution to Boltzmann equation \eqref{eq:Boltz}. We are interested in the yield at late times, $x \gg \xd$, where $\xd$ is the point at which dark matter ``detaches'' from the equilibrium curve. At late times we can make the approximation that $\Ychieq \approx 0$, and \eqref{eq:Boltz} reads
\beq \label{eq:ApproxBoltz}
\frac{d \Ychi}{dx}\approx-\frac{A}{x^{n+2}}\Ychi^2 + \frac{B}{x^{4-2p}},
\eeq
which is now in the form of a Riccati equation. Using the boundary condition $\Ychi(\xd) \approx \Ychieq(\xd)$ and by taking $\sqrt{\frac{A}{B}}(x_d)^{(\be-\al)/2}\Ychieq(\xd)\gg 1$, which holds for all scenarios considered in this work, we may solve this to find
\beq \label{eq:Yield}
Y_{\chi}(\infty) \approx  (\al+\be)^{\frac{\be-\al}{\al+\be}} \frac{ B^{\frac{\al}{\al+\be}} \Gamma\left(\frac{\be}{\al+\be}\right) I_{\frac{-\al}{\al+\be}}\left(\frac{2\sqrt{AB}}{(\al+\be)\xd^{(\al+\be)/2}}\right) }{ A^{\frac{\be}{\al+\be}} \Gamma\left(\frac{\al}{\al+\be}\right) I_{\frac{\al}{\al+\be}}\left(\frac{2\sqrt{AB}}{(\al+\be)\xd^{(\al+\be)/2}}\right)},
\eeq
where $\al=n+1$ and $\be=3-2p$. This solution is valid for $p<3/2$. In solving this an upper value of $x$ has been set to $\infty$, which in the cases we would like to study is an excellent approximation. However, if $p \geq 3/2$, $Y \rightarrow \infty$ as $x\rightarrow \infty$. As a result the upper limit for $x$ must be set to a finite value, which should be the value of $x$ at the present day. However, we are unaware of models that predict $p>3/2$ and we therefore do not consider this case.

The final part of the calculation is to find the point at which the yield departs from the equilibrium curve, $\xd$. In analogy with the standard freeze-out treatment \cite{Kolb:1994} we can determine $\xd$ by defining it as the point where $\Ychi (\xd) -\Ychieq(\xd) \approx c \Ychieq (\xd)$, where $c$ is a numerical constant of order unity. We can then use this relation with~\eqref{eq:ApproxBoltz} to find the recursive formula

\beq \label{eq:XdRec}
\begin{split}
\xd = & \log[A c(c+2)k]-\left(n+\frac{1}{2}\right)\log\left[\xd\left(1-\frac{2}{3\xd}\right)\right] \\&\hspace{3cm} -\log\left[\frac{1}{2}\left(1 + \sqrt{1+\frac{4Ac(c+2)B}{\left(1-\frac{2}{3\xd}\right)^2 \xd^{6+n-2p}}}\right)\right],
\end{split}
\eeq
where $k=(45/4\sqrt{2\pi^7})(g/g_*)$ and $g$ is the number of internal degrees of freedom of the dark matter state.
After one iteration we find
\beq
\begin{split} \label{eq:xd}
\xd \approx & \log[Ac(c+2)k]-\left(n+\frac{1}{2}\right)\log[Ac(c+2)k] \\& \hspace{3cm} -\log\left[\frac{1}{2}\left(1 + \sqrt{1+\frac{4Ac(c+2)B}{(\log[Ac(c+2)k])^{6+n-2p}}}\right)\right].
\end{split}
\eeq

The parameter $c$ can be fitted for best agreement with the numerical results but here we take $c(c+2) = n+1$, which gives sufficient agreement for our purposes. We note that if we take $B\rightarrow 0$ in~\eqref{eq:xd} the resulting expression for $\xd$ is the usual one for the freeze-out temperature in the standard freeze-out scenario, as one would expect \cite{Kolb:1994}. 

We can take two useful limits of the analytic solution presented in~\eqref{eq:Yield}. The first is where the product $AB$ is large. In this limit the solution reduces to the form
\beq \label{eq:highTDYield}
Y_{\chi}(\infty) \approx (\al+\be)^{\frac{\be-\al}{\al+\be}} \frac{ B^{\frac{\al}{\al+\be}} \Gamma\left(\frac{\be}{\al+\be}\right)}
{A^{\frac{\be}{\al+\be}}\Gamma\left(\frac{\al}{\al+\be}\right)}.
\eeq
The first thing to note is that this result is independent of $\xd$. Determining the freeze-out parameter $\xd$ from~\eqref{eq:xd} would suggest that the largest effect on $\xd$ from the ``$B$''-term correction comes when $B$ is largest. However, as is evident from~\eqref{eq:highTDYield}, this correction plays no role in determining the final yield. 

The second limit that we can take is where the contribution from defects is very small. In this limit the yield given in~\eqref{eq:highTDYield} reduces to the usual formula for standard freeze-out, see e.g. \cite{Kolb:1994}. 

\begin{figure}[t]
\centerline{\includegraphics[width=9cm]{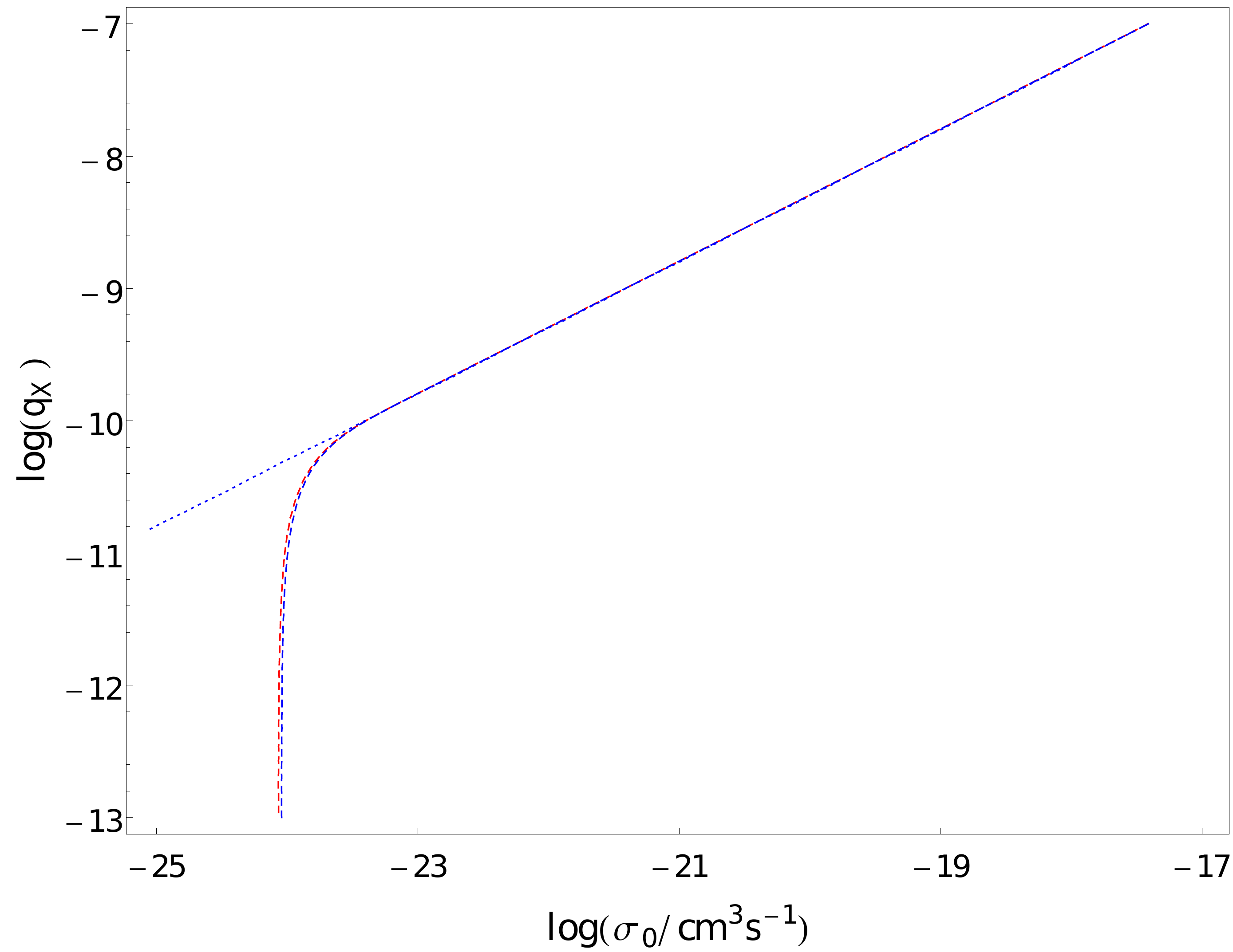}}
\caption{Comparison of numerical and analytic solutions. The curves represent the contours of constant yield fixed to the Planck-determined value. 
The red dashed line is derived from the numerically determined solution for the yield from the full Boltzmann equation~\eqref{eq:Boltz}, the blue dashed line repents the full analytic solution \eqref{eq:ApproxBoltz} and the straight blue dotted line represents the approximate solution~\eqref{eq:highTDYield}. All three are plotted for $(n, p)=(1, 1)$, with a $m_{\chi}=500\;$GeV, $\chimult=0.5$ and $\xref=1$.}
\label{fig:comp}
\end{figure}

In Figure~\ref{fig:comp} the full numerical solution (red dashed), the full analytic solution~\eqref{eq:Yield} (blue dashed) and the large $AB$ limit~\eqref{eq:highTDYield} (blue dotted) have been plotted for the case of $n=1, p=1$ for a $500$\;GeV dark matter mass. We can see that there is excellent agreement between the numerical and full analytic result. This agreement is repeated for all of our other scenarios.

In addition, by observing that the full analytic solution~\eqref{eq:Yield} reduces to the high defect approximation~\eqref{eq:highTDYield} when the Bessel functions are roughly equal (which is where their arguments are $\sim 2$) we can determine the following condition for when we can accurately describe the yield using \eqref{eq:highTDYield}
\bea
\label{eq:largedcon}
\Xprodpar\si_0 \gtrsim 10^{-38}\;\xd^{\al+\be}\;{\rm cm}^3\,{\rm s}^{-1}\;\left(\frac{500\;{\rm GeV}}{m_{\chi}}\right)\left(\frac{0.5}{\chimult}\right).
\eea
This condition proves to be reliable for all scenarios considered in this work.

Finally, we point out that one may think one could simply integrate the source term from the point at which the dark matter states would freeze-out in the usual scenario without the source term. This presumes annihilations will not be important post-freeze-out, whereas we have shown in this section and Section~\ref{sec:Numeric} that they are still active and play a crucial role in determining the final yield.

\section{Scenarios and constraints}\label{cons}

We would like to constrain the parameter space presented in Figure~\ref{fig:qxsig}. Before we mention model-dependent limits there is a dark matter model-independent limit that can be applied. It is clear that the dark matter annihilation cross section cannot be increased arbitrarily as at some point we will hit the unitarity limit. A formulation of this was derived in \cite{Griest:1989wd} and takes the form
\ben\label{unit}
\left< \si v\right> \leq \frac{4 (2n+1)\sqrt{\pi \xd}}{\mchi^2}.
\een
In arriving at this limit it is assumed that the annihilation cross section stays approximately constant throughout freeze-out \cite{Griest:1989wd}. Applying this limit to our scenario it is clear that it will lead to a different constraint on $\si_0$ for the two cases of $n=0$ and $n=1$ due to the different powers of $x$ in the expressions for an s-wave annihilation compared to p-wave.

\begin{figure}[t]
\begin{subfigure}{.5\textwidth}
  \hspace{-1cm}
  \includegraphics[width=8cm]{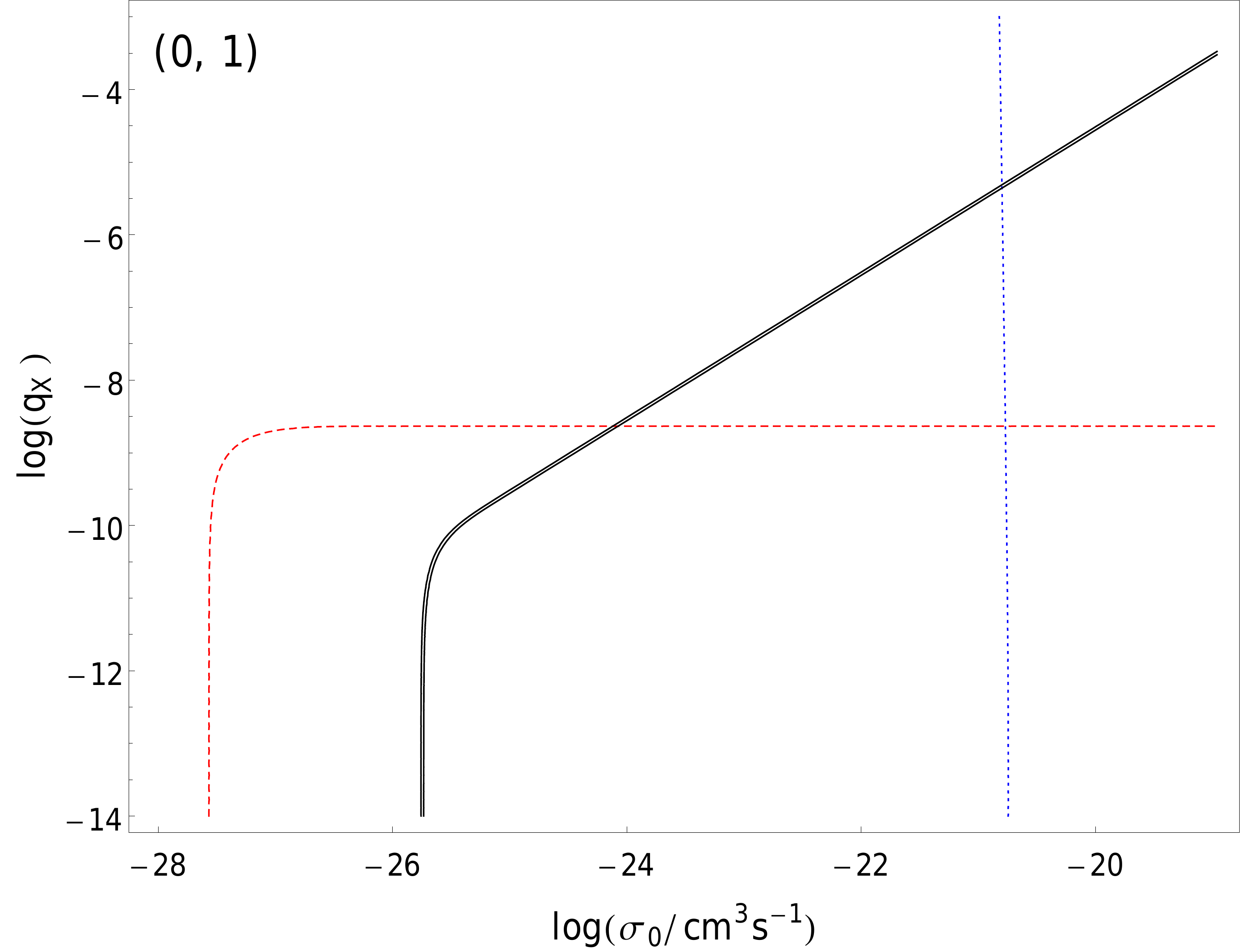}
  \caption{\phantom{    .}}
  \label{fig:LimitsN0P1}
\end{subfigure}%
\begin{subfigure}{.5\textwidth}
  \hspace{-0.0cm}
  \includegraphics[width=8cm]{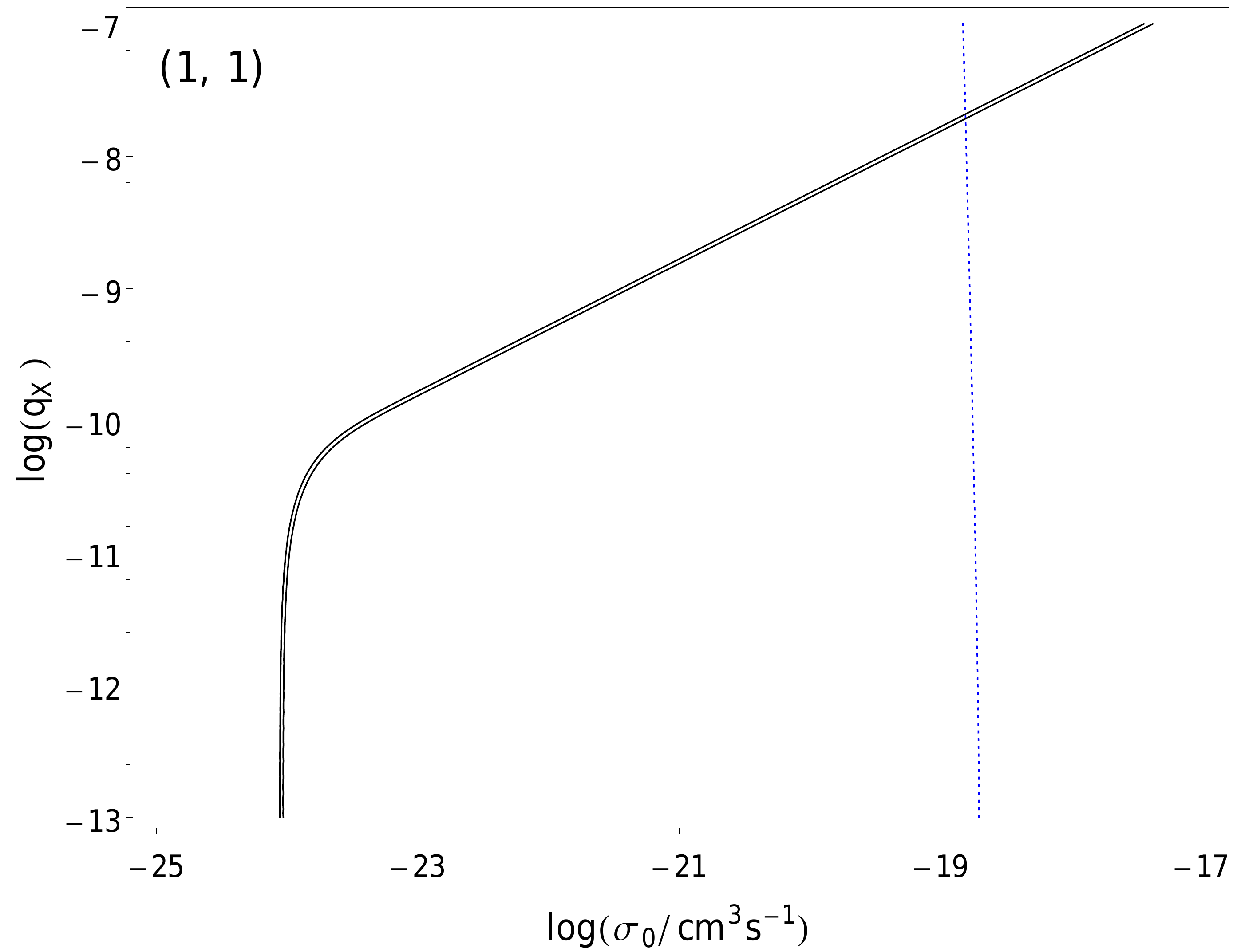}
  \caption{\hspace{-2.0cm}\phantom{    .}}
  \label{fig:LimitsN1P1}
\end{subfigure}%
\\\begin{subfigure}{.5\textwidth}
 \hspace{-1cm}
  \includegraphics[width=8cm]{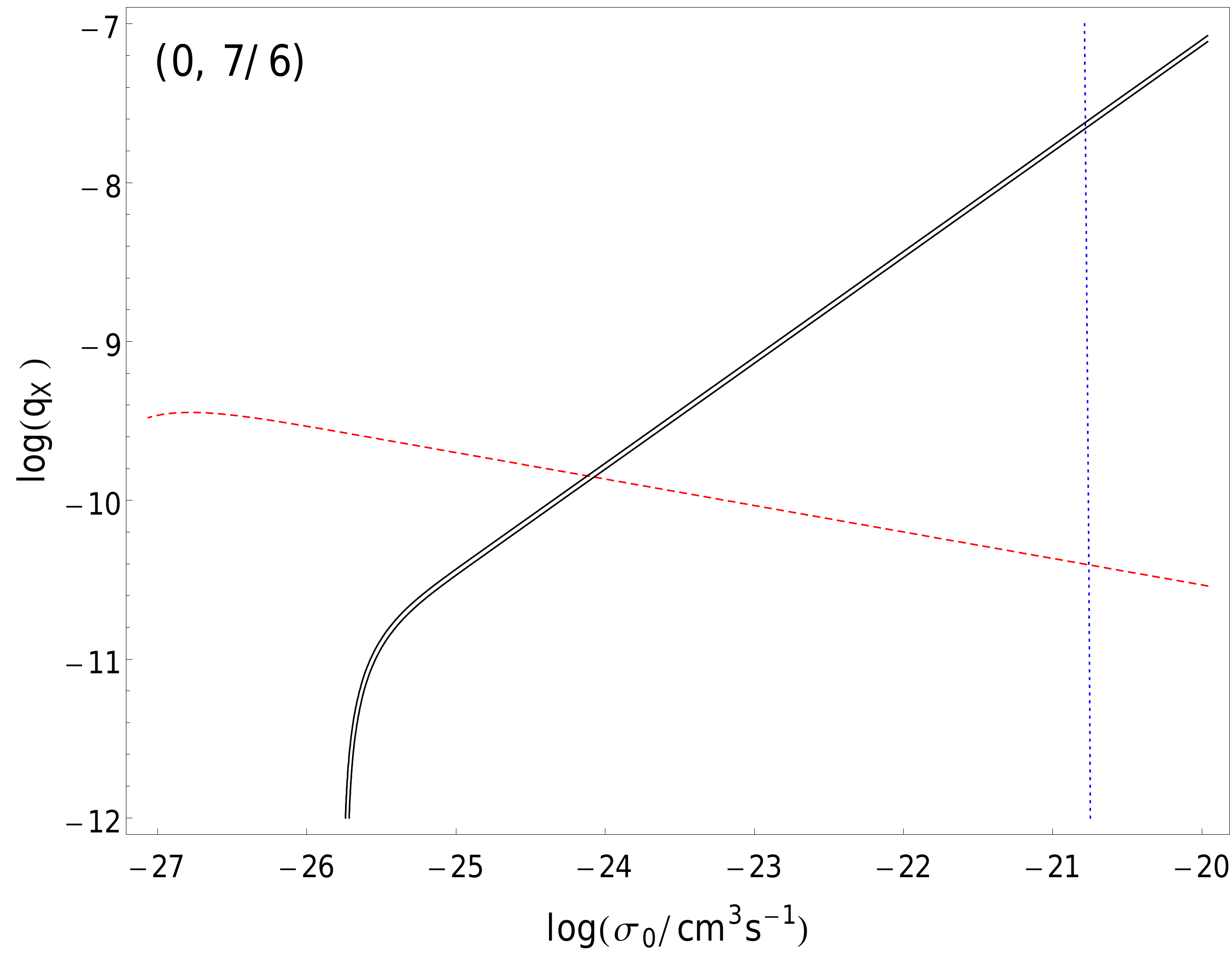}
  \caption{\phantom{    .}}
  \label{fig:LimitsN0P05}
\end{subfigure}%
\begin{subfigure}{.5\textwidth}
  \hspace{-0.0cm}
  \includegraphics[width=8cm]{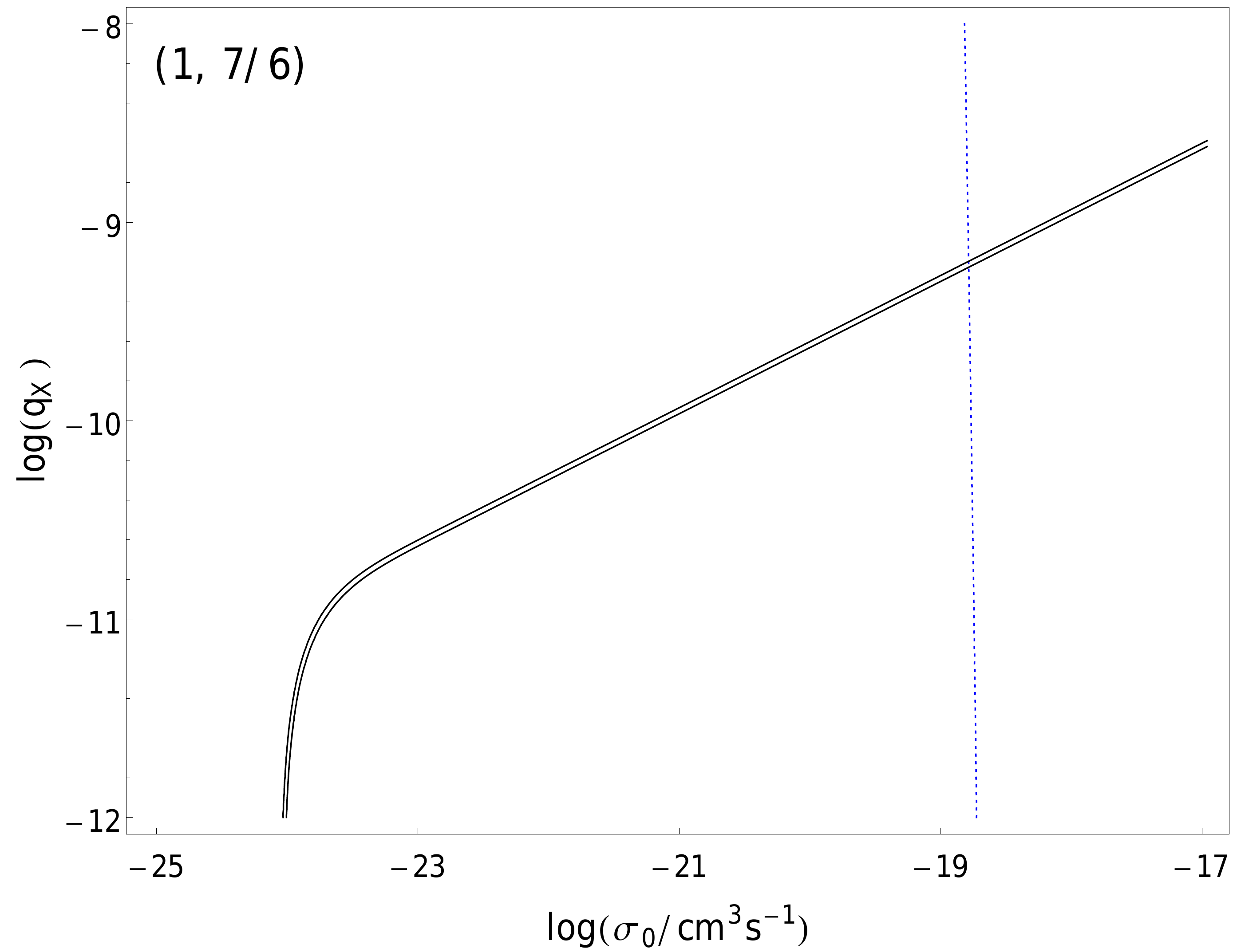}
  \caption{\hspace{-2.0cm}\phantom{     .}}
  \label{fig:LimitsN1P05}
\end{subfigure}%
\caption{Example constraints on the $\Xprodpar$ vs $\si_0$ parameter space. The solid black line is a constant yield contour, fixed to give the Planck value for the dark matter density. Also plotted is 
 the unitarity limit \cite{Griest:1989wd} (dotted blue line) and
 for the $n=0$ scenarios we have plotted an upper bound from the Fermi-LAT dwarf spheroidal galaxy limits \cite{drlica} (dashed red line). 
In the top left corner of the figures are given $(n,p)$, where $n$ is this orbital angular momentum of the annihilation, and $p$ is the parameter defined in~\eqref{eq:eninj}.}
\label{fig:nsnps}
\end{figure}

If Figure~\ref{fig:nsnps}, we have applied the unitarity limit in the ($\Xprodpar, \si_0$) plane for our four scenarios alongside a constant yield contour set to the experimentally measured value for a dark matter state with a $500\gev$ mass. The unitarity limits appear as near-vertical\footnote{The unitarity limit in the ($\Xprodpar, \si_0$) plane is not a perfectly vertical line due to a slight variation in $\xd$.}  lines (blue dashed) with everything to the right of the line ruled out. 

The unitarity bound is only determined by the mass of the dark matter state and whether it annihilates predominantly via s- or p-wave. It thus provides a very general and (dark matter) model-independent bound on $\Xprodpar$. The yield can be written using~\eqref{eq:highTDYield}, as the unitarity bounds correspond to $\Xprodpar$ and $\si_0$ values that satisfy~\eqref{eq:largedcon}, and as a result the limit on $\Xprodpar$ from unitarity can easily be found analytically as a function of the parameters of the scenario. The result is presented in Appendix~\ref{unitarity}.

In Figure~\ref{fig:nsnps}, an example of a dark matter model-dependent constraint has also been applied to give the plots some context. The example limit comes from the constraints arising as a result of searches for annihilations of dark matter states in the $\ga$-ray continuum from dwarf spheroidal satellite galaxies of the Milky Way (dSphs) with the Fermi-LAT experiment \cite{Ackermann:2011wa, drlica}. In particular we have used the preliminary 4 year Pass 7 analysis from Fermi-LAT presented in \cite{drlica}. In order to overlay these limits in the ($\Xprodpar, \si_0$) plane we need to choose a dark matter model. We are only concerned with giving an example of the interplay of these model-dependent indirect limits with our scenario and so we make a simplifying assumption that our dark matter states annihilate dominantly to a pair of $W$ bosons. In addition, we also assume that the dark matter annihilation is dominantly s-wave and we are therefore able to directly read off the resulting limit on $\langle \si v\rangle$ presented in \cite{drlica} for a dark matter mass of 500 GeV. 

The advantage of considering an s-wave annihilation is that $\langle \si v\rangle$ is velocity independent and will therefore remain constant from freeze-out to now. This means that we are able to directly apply constraints to $\langle \si v\rangle \approx \si_0$ from the indirect limits derived from the Fermi-Lat data in the ($\Xprodpar, \si_0$) plane. These limits are shown for the $n=0$ cases in Figures~\ref{fig:LimitsN0P1} and \ref{fig:LimitsN0P05} as red dashed lines.

We can, in principle, apply the Fermi-LAT limits to the $n=1$ cases as well but for a p-wave process $\langle \si v \rangle$ will be a velocity dependent quantity. In order to apply limits directly to $\si_0$ we would need to take account of this velocity dependence. This is a non-trivial exercise and well beyond the scope of this work. In addition, the mean velocity for dark matter particles in dwarf galaxies can be as low as a few $\text{km}\,\text{s}^{-1}$, which will lead to an extremely weak constraint on $\si_0$ at freeze-out, well above the limit derived from unitarity. Given this and the uncertainties associated with the velocity distributions of dark matter in dwarf galaxies we do not include these limits for the p-wave cases.

For the $n=0$ cases in Figures~\ref{fig:LimitsN0P1} and \ref{fig:LimitsN0P05} the shape of the Fermi-LAT limit can be understood in terms of the analytic solutions presented in~\eqref{eq:Yield} and~\eqref{eq:highTDYield}. For values of $\Xprodpar$ and $\si_0$ satisfying the condition in~\eqref{eq:largedcon}, the abundance of dark matter is set by~\eqref{eq:highTDYield}. In Figure~\ref{fig:LimitsN0P1}, with $(n, p)=(0, 1)$,  we have that $\al=\be$ and so~\eqref{eq:highTDYield} becomes $Y^{\al=\be}_{\chi}(\infty) \approx \sqrt{B/A}\propto \sqrt{\Xprodpar/\si_0}$. The predicted photon signal rate, call it $R$, from dark matter annihilations scales roughly as $R\propto \rho_{\rm DM}^2 \langle \si v \rangle \propto \Xprodpar$ and is therefore independent of the size of the cross section and is why we see a flat limit. For smaller values of $\Xprodpar$, we see the expected turnover of the indirect limit as it tends to what is expected from a freeze-out only scenario. 

For $(n, p)=(0, 7/6)$, shown in Figure~\ref{fig:LimitsN1P05}, $\al = 3\be/2$ and~\eqref{eq:highTDYield} becomes $Y^{\al=3\be/2}_{\chi}(\infty) \propto B^{3/5}/A^{2/5}\propto \Xprodpar^{3/5}/\si_0^{2/5}$ leading to a signal rate $R\propto \Xprodpar^{6/5}\si_0^{1/5}$. The resulting constraint will have some $\si_0$ dependence and the allowed value of $\Xprodpar$ decreases with increasing cross section (at the same time the dark matter abundance is also decreasing as we move to larger cross sections along the Fermi-LAT constraint line).

\begin{table}[htdp]
\begin{center}
\begin{tabular}{|l|l|l|l|}
\hline
$(n, p)$ & DM density & Unitarity & dSph $\ga$ emission  \\\hline
$(0, 1)$ & $\Xprodpar\lesssim 2.9\times 10^{-8}(\si_0^{23})$ & $\Xprodpar\lesssim 4.6\times 10^{-6}$ & $\Xprodpar\lesssim 2.3\times 10^{-9}$  \\
$(1, 1)$ & $\Xprodpar\lesssim 1.6\times 10^{-10}(\si_0^{23})^\frac12$ & $\Xprodpar\lesssim 2.0\times 10^{-8}$ & \hspace{1.2cm}-  \\
$(0, 7/6)$ & $\Xprodpar\lesssim 7.6\times 10^{-10}(\si_0^{23})^\frac23$ & $\Xprodpar\lesssim 2.3\times 10^{-8}$ & $\Xprodpar \lesssim1.4\times 10^{-10}$   \\
$(1, 7/6)$ & $\Xprodpar\lesssim 2.4\times 10^{-11}(\si_0^{23})^\frac13$ & $\Xprodpar\lesssim 6.1\times 10^{-10}$ & \hspace{1.2cm}- \\   
\hline
\end{tabular}
\caption{
Limits on the $X$ particle energy density injection rate parameter $\Xprodpar$, assuming that dark matter is derived from cosmic string decays.
The second column is the limit from the Planck value for the dark matter relic density, with $\si_0^{23} = \si_0/10^{-23}\, \cm^3\,
\text{s}^{-1}$, and applies when 
$\si_0\gg1.6\times 10^{-26}\;$cm$^3$s$^{-1}$ ($n=0$) or  
$\si_0\gg7\times 10^{-25}\;$cm$^3$s$^{-1}$ ($n=1$), 
the standard freeze-out values needed to generate the measured dark matter abundance. 
The other limits are obtained by combining the second column with the limits on the annihilation cross-section from unitarity \cite{Griest:1989wd} and Fermi-LAT bounds on $\ga$ emission in dSphs \cite{Ackermann:2011wa, drlica}. 
We have taken the dark matter multiplicity parameter $\chimult = 0.5$ (\ref{e:chimultDef}). An analytic formula giving the detailed parameter dependence of these limits is presented in Appendix~\ref{unitarity}.}
\label{tab:consqx}
\end{center}
\end{table}

A full summary of the limits on $\Xprodpar$ are displayed in Table~\ref{tab:consqx}. We have chosen as a default $\chimult=0.5$. We show an approximate expression for the limit on $q_X$ from the Planck value for the dark matter density,  for values of the annihilation cross-section larger than the values needed to generate the measured dark matter abundance in standard freeze-out. With the upper limits on the cross-section from 
unitarity and from Fermi-LAT bounds on $\ga$ emission in dSphs, we can derive upper limits on $\Xprodpar$ which are also shown. For a different value of $\chimult$, the limits in Table~\ref{tab:consqx} can simply be multiplied by a factor of $(0.5/\chimult)$. However, this must be done at constant dark matter mass as the limits depend non-trivially on $m_{\chi}$, see again \ Appendix~\ref{unitarity}. 

The limits coming from the dark matter search experiments have the potential to be much more constraining than the unitarity limit, but they are highly model-dependent.  Further sources of indirect constraints can be derived from data obtained from searches for dark matter in the Galactic centre \cite{Hooper:2012sr} for example and in a specific dark matter model, limits from direct detection may also be applied, however, the details of a fully model dependent analysis of the allowed parameter space for a particular dark matter model is beyond the scope of the present work. The aim of this section is to demonstrate the potential interplay between dark matter search limits and our parameter space. 

We note that even when these example indirect limits are applied, a significant and interesting viable parameter space still remains. With the source term it is possible to increase the size of the annihilation cross section beyond the standard freeze-out value, whilst still generating the correct relic abundance of dark matter. This may lead to a possible source of ``boost factors'' for indirect detection needed to explain possible anomalies in the positron fraction of cosmic rays \cite{Aguilar:2013qda}. In principle the size of the boost factor can be several orders of magnitude with a model-independent maximum coming from the unitarity limit. Of course, model dependent limits coming from both direct and indirect limits for the specific dark matter candidate need to be applied but even with the specific indirect limit applied above we see that we still have nearly two orders of magnitude in terms of a possible boost factor for both (0, 1) and (0, 7/6) cases. 

The limits on $\Xprodpar$ can be used to constrain specific \TD\ scenarios, including those presented in Appendix~\ref{s:EvoStr}. Using~\eqref{qxvdft} and \eqref{qxguce}, we find limits on the cosmic string tension parameter $G\mu$ for all four scenarios. The results are shown in Table~\ref{vdlims}.
Again, we have chosen as a default $\chimult=0.5$, and to calculate the limits 
for a different value of $\chimult$, the values in Table~\ref{tab:consqx} can again be multiplied by a factor of $(0.5/\chimult)$.
We note that the limits are generically well below $G\mu \sim 10^{-7}$, corresponding to the scale of Grand Unification or inflation and current Cosmic Microwave Background limits \cite{Ade:2013xla}. From (\ref{qxguce}) one sees that, given that the bound on $q_X$ is inversely proportional to $ r_{\chi}$, upper bounds on $G\mu$ are also inversely proportional to $r_{\chi}$. Hence limits recede with increasing average $X$ state energy $\EbarX$, and do not significantly constrain $G\mu$ if $\EbarX$ is of order the string mass scale $\vd$.

\begin{table}[t]
\begin{center}
\begin{tabular}{|l|l|l|l|}
\hline
$(n, p)$ & DM density & Unitarity & dSph $\ga$ emission \\\hline
$(0, 1)$ & $G\mu< 1.1\times 10^{-10}(\si_0^{23})\paramFT^{-1}$ & $G\mu< 1.7\times 10^{-8}\paramFT^{-1}$ & $G\mu< 8.7\times 10^{-12}\paramFT^{-1}$  \\
$(1, 1)$ & $G\mu< 6.0\times 10^{-13}(\si_0^{23})^\frac12\paramFT^{-1}$ & $G\mu< 7.5\times 10^{-11}\paramFT^{-1}$ & \hspace{1.8cm}-   \\
$(0, 7/6)$ & $G\mu< 1.1\times 10^{-14}(\si_0^{23})^\frac23\paramCE^{-1}$ & $G\mu < 3.4\times 10^{-13} \paramCE^{-1}$ & $G\mu< 2.1\times 10^{-15} \paramCE^{-1}$\\
$(1, 7/6)$ & $G\mu< 3.6\times 10^{-16}(\si_0^{23})^\frac13\paramCE^{-1}$ & $G\mu < 9.1\times 10^{-15} \paramCE^{-1}$ &\hspace{1.8cm}-\\
\hline
\end{tabular}
\caption{Limits on the cosmic string tension parameter $G\mu$ in the two cosmic string scenarios \FTscen\ and \CAscen, described in Appendix \ref{s:RadStr}, assuming that dark matter is derived from cosmic string decays. The limits are from unitarity of the annihilation cross-section \cite{Griest:1989wd}, and Fermi-LAT bounds on $\ga$ emission in dSphs \cite{Ackermann:2011wa, drlica}.
We have taken the dark matter multiplicity parameter $\chimult = 0.5$ (\ref{e:chimultDef}), and the O(1) parameter combinations $\paramFT$ and $\paramCE$ are defined in (\ref{e:paramFTDef}) and (\ref{e:paramCEDef}).
\label{vdlims}}
\end{center}
\end{table}%

\section{Conclusions}

In this paper we have studied dark matter production in top-down or topological defect  (\TD) models of particle production in the early universe, with two different cosmic string scenarios as specific examples. 
\TD\ models introduce a source term for the production of dark matter with a characteristic power-law dependence on time. The source term can be simply parametrised by an amplitude and a power law index.
We have found an analytic formula (\ref{eq:Yield}) for the yield which is applicable to all such scenarios, attaining a particularly simple form (\ref{eq:highTDYield}) in the limit that most of the dark matter is derived from the \TD\ source.  
The effect of the source is to allow higher relic densities for values of the annihilation cross-section above that required by the standard freeze-out calculation. Hence the \TD\ scenario is a potential source of boost factors that can be used to increase the predicted signal rates in indirect detection.

In two \TD\ scenarios representing cosmic strings, and considering both s- and p-wave annihilation, we have examined the limits in the plane of the energy injection rate $\Xprodpar$ and the annihilation cross-section parameter $\si_0$, displayed in Figure \ref{fig:nsnps}.  We have derived an upper bound on $\Xprodpar$ as a function of $\si_0$ from the dark matter relic density. As this bound is an increasing function of $\si_0$, the upper bound on $\si_0$ from unitarity gives an upper bound on $\Xprodpar$ in a dark matter model-independent way. 
We have also examined limits from searches for $\ga$ emission from dwarf spheroidal galaxies. While this is a model-dependent question,  we derive representative limits under the assumption that the dark matter states annihilate dominantly to a pair of $W$ bosons.
The limits are displayed in Table \ref{tab:consqx}.

The tightest model-independent constraint on $\Xprodpar$ comes from the $(n, p)=(1, 7/6)$ scenario, corresponding to dark matter states that annihilate dominantly via a p-wave process and a source term that represents cosmic strings in the cusp emission scenario. The limit of $\Xprodpar < 6.1\times 10^{-10}$ derived from the unitarity of the dark matter annihilation rate shows that dark matter physics can play an interesting role in constraining the properties of topological defects. 

In string scenarios, the bound on $\Xprodpar$ translates into a bound on the string tension parameter $G\mu$. In all cases studied, the bounds on $G\mu$ (displayed in Table \ref{vdlims}) were well below the Grand Unified or inflation scale ($G\mu \sim 10^{-7}$).  The only way to significantly weaken the bounds would be to suppose that the average energy of the $X$ particles emitted by the string was much greater than $1\, \TeV$, hence greatly reducing the parameter $\chimult$.

The logical next step would be to study specific models of dark matter and cross correlate the analysis in this work with limits coming from direct detection, collider physics and cosmic rays. This could be with a view to establish working models of dark matter with an altered prediction for indirect rates compared to standard freeze-out or it could be with a view to limit further the properties of topological defects. We postpone this detailed analysis to future work.

\acknowledgments

This research was supported in part by the Science and Technology Facilities Council [grant numbers ST/J000485/1 and ST/J000477/1]. SMW thanks the Higher Education Funding Council for England and the Science and Technology Facilities Council for financial support under the SEPNet Initiative. MH thanks Herondy Mota for helpful conversations.

\appendix

\section{Particle radiation from cosmic strings}

\label{s:RadStr}

\subsection{Evolution of strings}

\label{s:EvoStr}

Strings can be created in a phase transition if the symmetry-breaking produces a multiply-connected vacuum manifold \cite{Kibble:1976sj}.  Denoting the vacuum expectation value of the symmetry-breaking scalar field $\vd$, the mass per unit length of the strings $\mu$ is proportional to $\vd^2$. 
The strings are in the form of three-dimensional random walks with a step length $\xi_c$ determined by the transition rate \cite{Kibble:1976sj, Zurek:1996sj}. At the time of the transition $t_c$, the majority of the length is in one string which is a large as the universe (or infinite in an infinite universe). The rest is in closed loops with a scale-free length distribution
$n(\ell,t_c) \sim \ell^{-\frac52}$.
It is conventional to call strings are those whose length is greater than the horizon length ``long'' or ``infinite'', and those whose length is less than the horizon length are called loops.

It is also conventional to introduce a parameter with the dimensions of length $\xi$, defined from the energy density of long strings $\rho_\infty$ and the mass per unit length $\mu$, as 
\ben
\xi = \sqrt{\mu/\rho_\infty}.
\een
It can be interpreted as a combination of the average curvature radius and the average separation of long string segments.  

The strings evolve by straightening under their own tension, subject to Hubble damping and friction from the cosmic fluid \cite{Kibble:1980mv}.  Friction is important until $\Tfree \sim \vd^2/\mpl$, or $\Tfree \sim (G\mu/10^{-16})^\half \, \TeV$. Subsequently, strings evolve under Hubble damping only, and enter the scaling regime during which 
\ben
\xi \propto t.
\een  
The long string network can lose energy by self-intersection and the formation of loops.  During the scaling era, the loop length distribution remains a topic of active debate \cite{Hindmarsh:2008dw,BlancoPillado:2011dq}. It seems that there are several scales involved, including the curvature radius at the end of the friction-dominated era $\xifree$. 

In the scaling era loops are freely oscillating, and can lose energy via a variety of channels.
In the original cosmic string scenario, perturbative particle production from an oscillating string loop was shown to be negligible \cite{Vachaspati:1984yi,Srednicki:1986xg}, and the most important source of energy loss was argued to be gravitational radiation for sufficiently large loops \cite{Vilenkin:1981bx}.  However, direct numerical simulations of field theories with strings showed that there was an additional, non-perturbative, source of particle production in the form of classical field radiation \cite{Vincent:1997cx,Hindmarsh:2008dw}. In either case the loop length $\ell$ shrinks at a constant rate
\ben
\dot \ell = -\be,
\label{e:LooDec}
\een
with $\be \sim 1$ in the case of non-perturbative particle production \cite{Hindmarsh:2008dw} and $\be = \Ga G\mu$ in the case of gravitational radiation. Here, $\Ga$ an efficiency parameter which is typically O(100) \cite{Vachaspati:1984gt}.  
The lifetime of a loop with length $\ell$ is therefore $\ell/\be$.

We consider two different string scenarios. In the first, motivated by direct numerical simulation of the field theory (\FTscen) 
\cite{Vincent:1997cx,Bevis:2006mj,Hindmarsh:2008dw}, it is assumed that particle production dominates at all times.  In the second, it is assumed that loops of string with large enough average curvature radii obey the Nambu-Goto equations, and that the dominant energy loss is into gravitational radiation \cite{VilShe94,Hindmarsh:1994re}.  There is also a source of particle production in the form of cusp emission (\CAscen). This can happen in two ways: strings double back on themselves and annihilate \cite{Brandenberger:1986vj,BlancoPillado:1998bv}, and the string can also be a classical source of massive scalar radiation, strongly beamed from the cusps \cite{Damour:1996pv,Vachaspati:2009kq}. 
In both cases the length decreases as \cite{BlancoPillado:1998bv,Vachaspati:2009kq} (\CAscen) 
\ben
\label{e:CEPow}
\dot \ell = -\beCE {\frac{1}{\sqrt{\mX\ell}}},
\een
where $\mX$ is the mass scale of the emitted $X$ particles, which is also the inverse string width.  
Numerical factors and couplings are gathered into a constant $\beCE$.
  
Cusp emission is subdominant for loops larger than a critical size $\ellCE = (\be_c/\be)^2 \mX^{-1}$.  The earliest time that such loops can be formed is 
\ben
\label{e:tCEDef}
\tCE = (\beCE/\be)^2 (\be\mX)^{-1},
\een 
and so cusp emission is more important for $t < \tCE$, where $\tCE$ has the value
\bea
\tCE 
&\simeq&  6.6 \times 10^{-13} \,\frac{\beCE^2}{\Ga_{100}^3}\left( \frac{10^{-7}}{G\mu} \right)^3 \left( \frac{\TeV}{\mX} \right)  \, \text{s}
\label{e:tCEvals}
\eea
The critical time for cusp emission corresponds to a temperature
\ben
\TCE = \left( \frac{45}{2\pi^2\geff} \right)^\frac14 \frac{\be^\frac32}{\beCE} \sqrt{\mpl\mX}.
\een
In terms of our reference parameters, 
\ben
\label{e:TCEval}
\TCE \simeq 600 \left( \frac{100}{\geff} \right)^\frac14 \frac{\Ga_{100}^\frac32}{\beCE} \left( \frac{G\mu}{10^{-7}} \right)^\frac32 
\left(\frac{\mX}{\TeV}\right)^\half\; \GeV.
\een
Hence we must use (\ref{e:CEPow}) to calculate the loop distribution at the dark matter reference temperature $\Tref = 500\,\GeV$ if
$G\mu \lesssim 10^{-7}$, corresponding to an upper limit on the symmetry-breaking scale $\vd \lesssim 10^{15}\, \GeV$.

\subsection{\FTscen\ scenario }

We denote the defect density parameter $\Omd = \rhod/\rho$, where $\rho$ is the total energy density, and their average equation of state parameter $\eosd$. With $w$ as the total average equation of state parameter, and $H$ the Hubble parameter, covariant energy conservation can be used to show that the energy density injection rate into other species from string decay is 
\ben
\frac{Q}{\rho H} = 3(w-\eosd)\Omd.
\een
At late times, cosmic strings enter a self-similar scaling evolution, with $\rhod = \mu/\xi^2$ and 
\ben
\Omd = \frac{8\pi G\mu}{3\xi^2 H^2},
\een
which is constant. 
Numerical simulations show that, for Abelian Higgs strings, $\xi \simeq 0.25 d_h$ in both radiation and matter eras, where $d_h$ is the horizon distance \cite{Bevis:2006mj,Bevis:2010gj}, and $\eosd$ is small and negative, but must satisfy $\eosd \gtrsim -1/3$ \cite{VilShe94}.
Hence, the $X$ particle energy density injection rate parameter in the radiation era is 
\ben\label{qxvdft}
\Xprodpar = \frac{Q\fX}{\rho H} \simeq 
\displaystyle 2.6\times 10^{-5} \fX \left(\frac{1-3\eosd}{2}\right)\left( \frac{0.25}{\xi H}\right)^2 \left( \frac{G\mu}{10^{-7}} \right).
\een
For convenience when discussing bounds, we will define an O(1) parameter
\ben
\label{e:paramFTDef}
\paramFT =  \fX \left(\frac{1-3\eosd}{2}\right)\left( \frac{0.25}{\xi H}\right)^2,
\een 
such that $\Xprodpar \propto \paramFT G\mu$.
Writing $\mu = 2\pi \vd^2 \muPar$, where $\muPar$ is an O(1) function with a weak dependence on couplings \cite{Hindmarsh:1994re}, we see that $G\mu = 10^{-7}$ corresponds to $\vd \simeq 1.5 \times 10^{15} \ep^{-\half}\, \GeV$.

\subsection{\CAscen\ scenario}

In this scenario, the loop distribution function has the scale-free form in the radiation era 
\ben
\label{e:LooLenDis}
n(\ell,t) = \frac{\nu}{t^\frac{3}{2}\ell_i^\frac{5}{2}}
\een
where $\ell_i$ is the average size of of a loop of size $\ell$ when it is formed, at time $t_i$,  and $\nu$ is a dimensionless O(1) parameter \cite{VilShe94,Hindmarsh:1994re,BlancoPillado:2011dq}.  
For $t\gg \tCE$, the loops radiate according to (\ref{e:LooDec}) and shrink so that 
\ben
\ell_i \simeq \ell + \be t,
\een
where we can neglect $t_i$ in comparison to $t$ as loops are long-lived ($\be^{-1} \gg 1$).
For $t \ll \tCE$, the loops radiate according to (\ref{e:CEPow}) and shrink so that 
\ben
\ell_i = \ellCE \left[ \left(\frac{\ell}{\ellCE}\right)^\frac32 + \frac32 \frac{t}{\tCE} \right]^\frac23.
\een
The energy density injection rate from cusp annihilation is 
\ben
\Qcusp  = \int_0^\infty d\ell \be_c \mu {\frac{1}{\sqrt{\mX\ell}}} n(\ell,t).
\een
Hence, the radiation era energy density injection rate is 
\ben
\Qcusp^\text{(r)}  = \frac{\nu\be_c}{\be^\half} \frac{ \mu}{t^3} C\left(\frac{t}{\tCE}\right),
\een
where (on defining $\tau = {t}/{\tCE}$)
\ben
C(\tau) \simeq
\left\{ 
	\ba{cc}
		3.0 \, \tau^\frac16, & \text{for} \; \tau \ll 1  \cr
		\frac{4}{3} \tau^{-\frac12}, & \text{for} \;  \tau \gg 1  \cr
	\ea
\right.
\label{e:QcuspNG}
\een
and we see that  $p = 7/6$  (for $t < \tCE$) or  $p = 1/2$ (for $t > \tCE$) , where $p$ is defined in (\ref{eq:eninj}).

The the $X$ particle energy density injection rate parameter can now be written 
\bea
\Xprodpar = \left. \frac{\Qcusp^\text{(r)}\fX}{\rho H} \right|_{\Tref} &\simeq & 
\frac{64\pi}{3}\frac{\be_c\nu\fX}{\Ga^\half} (G\mu)^\half C(\tref/\tCE), 
\eea
Now, using (\ref{e:TCEval}) one finds
\ben
\tau_\ka = \frac{\tref}{\tCE} \simeq 1.4 \left( \frac{100}{\geff} \right)^\frac12 
	\frac{\Ga_{100}^3}{\beCE^2} \left( \frac{G\mu}{10^{-7}} \right)^3 \left(\frac{\mX}{\TeV}\right)^{\half}\left(\frac{500\,\GeV}{\Tref}\right)^2.
\een
Hence
\ben\label{qxguce}
\Xprodpar \simeq 
\left\{ 
	\ba{cc}
		\displaystyle 6.8  \times 10^{-3} \, \left(\nu \beCE^\frac23 \fX\right) \left( \frac{100}{\geff} \right)^\frac{1}{12} \left( \frac{G\mu}{10^{-7}} \right) \left(\frac{\mX}{\TeV}\right)^{\frac16} \left(\frac{500\,\GeV}{\Tref}\right)^{\frac13}   
		& \text{for} \; G\mu \ll 10^{-7} \Dref \cr
		\displaystyle  2.4  \times 10^{-3} \, \left(\frac{\nu \beCE^2 \fX}{\Ga_{100}^2}\right) \left( \frac{\geff}{100} \right)^\frac{1}{4} \left( \frac{10^{-7}}{G\mu} \right) \left(\frac{\TeV}{\mX}\right)^{\frac12} \left(\frac{\Tref}{500\,\GeV}\right)
		& \text{for} \;  G\mu \gg 10^{-7} \Dref \cr
	\ea
\right.
\een
where the parameter $\Dref$ which divides ``large'' and ``small'' $G\mu$ is given by 
\ben
\Dref \simeq 0.89 \left( \frac{\geff}{100} \right)^\frac16 
	\frac{\beCE^\frac23}{\Ga_{100}} \left(\frac{\TeV}{\mX}\right)^{\frac16}\left(\frac{\Tref}{500\,\GeV}\right)^\frac23.
\een
Interestingly, $\Xprodpar$ reaches a maximum at around $G\mu \simeq 10^{-7}$ for our model parameters $\mX = 1\,\TeV$ and $\Tref = 500\,\GeV$.  Values of $G\mu$ significantly larger than $10^{-7}$ are ruled out by Cosmic Microwave Background data \cite{Ade:2013xla}, which means that only the low $G\mu$ form is relevant.
Again, for convenience when discussing bounds, we will define an O(1) parameter
\ben
\label{e:paramCEDef}
\paramCE =  \left(\nu \beCE^\frac23 \fX\right) \left( \frac{100}{\geff} \right)^\frac{1}{12}\left(\frac{\mX}{\TeV}\right)^{\frac16} \left(\frac{500\,\GeV}{\Tref}\right)^{\frac13},
\een 
such that $\Xprodpar \propto \paramFT G\mu$ for low $G\mu$.

One may be concerned that loops are still friction dominated at the important temperature $\Td \simeq \mchi/25$ where the relic dark matter is being produced.  For $\mchi = 500\;\GeV$ this is true only for $G\mu \lesssim 10^{-19}$, in a region well below our lowest bound in Table \ref{vdlims}. Hence, for the string tensions affected by our bounds, we can be sure that the friction-dominated period finishes well before freeze-out.

\section{Analytic forms for the limits on $\Xprodpar$}\label{unitarity}

In Table~\ref{tab:consqx}, the limits on $\Xprodpar$ are quoted for a number of constraints in our example scenario with a 500$\gev$ dark matter state. It is worth noting that provided condition~\eqref{eq:largedcon} is satisfied, we can use~\eqref{eq:highTDYield} to determine an analytic form for the limit on $\Xprodpar$ as a function the dark matter annihilation parameter $\si_0$. The result reads 

\beq
\Xprodpar \leq \frac{4}{3} \frac{1}{\chimult} (\al +\be)^{\frac{\al -\be}{\al}} \left(\sqrt{\frac{\pi \geff}{45}} M_{pl}m_{\chi}\; \si_0\right)^{\be/\al}\left[ \frac{\Ga(\frac{\al}{\al +\be})}{\Ga(\frac{\be}{\al +\be})}\frac{\Om_{\chi} \rho_c}{s_0 m_{\chi}}\right]^{\frac{\al +\be}{\al}}\hspace{-7mm}, \label{eq:sig0}
\eeq
where $\rho_c$ and $s_0$ are the critical energy density and entropy density respectively evaluated today, $\al=n+1$ and $\be=3-2p$, and we have set $\xref=1$ as before. Evaluating this further by setting the relic abundance to the measured value and setting other parameters to typical values we find

\begin{multline}
\Xprodpar \lesssim 2\times 10^{-8}\left(\frac{\al +\be}{10^4}\right)^{\frac{\al -\be}{\al}} \left(\frac{\Ga(\frac{\al}{\al +\be})}{\Ga(\frac{\be}{\al +\be})}\right)^{\frac{\al +\be}{\al}} \left(\frac{0.5}{\chimult}\right)\left(\frac{500\gev}{m_{\chi}}\right)\\ \times \left(\frac{\sqrt{\geff}}{10}\right)^{\frac{\be}{\al}}\left(\si_0^{23}\right)^{\frac{\be}{\al}}\left(\frac{\Om_{\chi}h^2}{0.1186}\right)^{\frac{\al+\be}{\be}}\hspace{-7mm},
\end{multline}
where $\si^{23}_0=\si_0/10^{-23}\mbox{cm}^3\mbox{s}^{-1}$. The result can be used to derive limits on $\Xprodpar$ from bounds on the dark matter annihilation cross section coming from model-dependent limits, for example the Fermi-LAT measurements of $\ga$ emission in dSphs \cite{Ackermann:2011wa, drlica} or if we would like to find the value of $\Xprodpar$ needed to generate the correct relic abundance for a given annihilation cross section.

We can further manipulate this expression to derive an analytic form for the limit on $\Xprodpar$ coming from the unitarity of the dark matter annihilation cross section. Applying the unitarity constraint in Eq.~\ref{unit} to $\si_0$ we have that
\begin{multline}
\Xprodpar \leq 2\times 10^{-12} \left( 4\times10^{5}\right)^{\frac{\be}{\al}}(\al +\be)^{\frac{\al -\be}{\al}}(2\al-1)^{\frac{\be}{\al}}\left(\frac{\Ga(\frac{\al}{\al +\be})}{\Ga(\frac{\be}{\al +\be})}\right)^{\frac{\al +\be}{\al}}x_d^{(2\al-1)\frac{\be}{2\al}}  \left(\frac{0.5}{\chimult}\right) \\
\times\left(\frac{500\gev}{m_{\chi}}\right)^{\frac{2\be+\al}{\al}}\left(\frac{\sqrt{\geff}}{10}\right)^{\frac{\be}{\al}}\left(\frac{\Om_{\chi}h^2}{0.1186}\right)^{\frac{\al+\be}{\be}}\hspace{-6mm}.
\end{multline}
To accurately determine the resulting numerical limit for a specific $(n, p)$, the value of $\xd$ needs to be evaluated from Eq~(\ref{eq:XdRec}).

\bibliography{Bibliography}
\bibliographystyle{JHEP}

\end{document}